\documentclass[a4,11pt]{article}

\usepackage{amsthm}
\usepackage{amsmath}
\usepackage{amsfonts}
\usepackage{amssymb}

\usepackage{bm}
\usepackage{gensymb}
\usepackage{mathtools}
\usepackage{xcolor}
\usepackage{authblk}

\usepackage{algorithm}
\usepackage{algorithmic}
\usepackage{hyperref}
\begin{document}

\newcommand{\fixme}[1]{ { \bf \color{red}FIX ME \color{black} #1 } }
\newcommand*{\affaddr}[1]{#1}
\newcommand*{\affmark}[1][*]{\textsuperscript{#1}}
\providecommand{\e}[1]{\ensuremath{\times 10^{#1}}}
\providecommand{\keywords}[1]{\textbf{\textit{ \\ Keywords:}} #1}
\providecommand{\acknowledgements}[1]{\textbf{\\ Acknowledgements:} #1}

\title{Simulation of hydro-mechanically coupled processes in rough rock fractures using an immersed boundary method and variational transfer operators}

\author[1]{Cyrill von Planta}
\author[3,4]{Daniel Vogler}
\author[3]{Xiaoqing Chen}
\author[1,2]{Maria G.C. Nestola}
\author[2]{Martin O. Saar}
\author[1]{Rolf Krause}

\affil[1]{Institute of Computational Science, Universit\`{a} della Svizzera italiana, Lugano,  6900, Switzerland}
\affil[2]{Center for Computational Medicine and Cardiology CCMC, Lugano, 6900,Switzerland}
\affil[3]{ETH Zurich, Institute of Geophysics, 8092 Zurich, Switzerland}
\affil[4]{ETH Zurich, Institute of Process Engineering, 8092 Zurich, Switzerland}

\date{Preprint, accepted version}
\maketitle

\vspace{1cm}
\begin{abstract}
Hydro-mechanical processes in rough fractures are highly non-linear and govern productivity and associated risks in a wide range of reservoir engineering problems. To enable high-resolution simulations of hydro-mechanical processes in fractures, we present an adaptation of an immersed boundary method to compute fluid flow between rough fracture surfaces. The solid domain is immersed into the fluid domain and both domains are coupled by means of variational volumetric transfer operators. The transfer operators implicitly resolve the boundary between the solid and the fluid, which simplifies the setup of fracture simulations with complex surfaces. It is possible to choose different formulations and discretization schemes for each subproblem and it is not necessary to remesh the fluid grid.
We use benchmark problems and real fracture geometries to demonstrate the following capabilities of the presented approach: (1) Resolving the boundary of the rough fracture surface in the fluid; (2) Capturing fluid flow field changes in a fracture which closes under increasing normal load; and (3) Simulating the opening of a fracture due to increased fluid pressure.
\end{abstract}
\section{Introduction}
Hydro-mechanical (HM) processes in discontinuities such as rock fractures play an important role in a number of reservoir engineering applications, for example enhanced geothermal systems, oil- and gas recovery or CO$_2$ sequestration. They govern productivity and thereby economical feasibility of reservoir operations \cite{tester_2006,mcclure_2014b,amann_2018}. 
Global reservoir permeability is governed by the hydraulic properties of single fractures as well as their degree of connectedness and geometric configuration in a fracture network \cite{cacas_1990,zimmerman_1991,rutqvist_2003,dreuzy_2012,hobe_2018}. 
The hydraulic properties of a single fracture are subject to constant change by various physical processes, which are often tightly coupled and highly non-linear. Given the impact of fracture hydraulics on resulting flow rates, HM processes in fractures are an actively researched topic. In the following text we exclusively focus on HM processes in single fractures.

For the mechanical behavior of a fracture, the highly variable fracture surface topographies yield a complex distribution of contact area and a non-linear mechanical response to normal or shear loading \cite{bandis_1983,pyraknolte_2000,jiang_2006,matsuki_2008,watanabe_2008,tatone_2015}. 
Mechanical loading normal to the fracture surface results in convergent fracture closure behavior with increasing load, as more and more area of the fracture comes into contact and local compressive stresses at contact areas increase \cite{bandis_1983,raven_1985,matsuki_2008,zangerl_2008,vogler_2016,vogler_2017,vogler_2018,kling_2018}. 

For the hydraulic behavior of fractures, rough fracture surface topographies result in highly heterogeneous aperture fields, which characterize the mechanical gap between two fracture surfaces \cite{zimmerman_1991,zimmerman_1996}. 
These heterogeneous aperture fields are naturally subject to change with mechanical loading and yield complex fluid flow patterns through fractures \cite{tsang_1984,brown_1987,tsang_1987,watanabe_2008,nemoto_2009,vogler_2018}. 
While an increase in mechanical loading leads to fracture closure, this can also result in elevated fluid pressures, as larger fluid pressure gradients are required to sustain fluid flow rates through closing fractures \cite{vogler_2016}. 
If the fluid pressure in the fracture increases sufficiently, it can counteract compressive mechanical load and even cause fractures to open.

Given the small scale surface topography and aperture variations, as well as the tightly coupled physical processes, simulation of HM phenomena can become computationally very expensive.  
Therefore, most numerical and analytical studies rely on simplifying assumptions when studying HM processes \cite{barton_1985,nemoto_2009,mcclure_2014a,figueiredo_2015b,GWG+15,gan_2016}. 
Commonly used assumptions are Darcy or lubricated flow, global or local parallel plate models for fluid flow in rock fractures, where averaged hydraulic and mechanical parameters are used globally or locally (i.e., on smaller surface patches) to capture the non-linear behavior of HM processes in rock fractures. 
The number of studies that model HM processes on real rough rock geometries, however, is small. The aim of this study is to present an approach to model HM processes in complex three dimensional geometries without the need to simplify them or the physical relations.

To this end, we use here ideas of the immersed boundary (IB) method which can be considered as a representative of the wider class of fictitious domain (FD) methods. The main feature of FD methods is to substitute the solution of a PDE on a complex domain by embedding or immersing it in a simpler domain on which it is easier to solve the problem. It is outside the scope of this article to give an in-depth review of the intertwined history of FD and IB methods. To this end we refer the reader to \cite{Hym52,Sau63,GPH+99,Baa01,Pes02,BG03,LLF+06,HGC+14} and the references therein.

 An FD method with fluid structure interaction (FSI) to simulate particulate flow was introduced by \cite{GPH+99}. There, the interaction is enforced with Lagrange multipliers distributed over each particle to match the velocities of the fluid and the moving rigid bodies. This method was later extended by \cite{Yu05} to deal with flexible bodies. \cite{Baa01} proposed a fictitious domain/mortar element method to impose the continuity of solid and fluid velocities along the solid-fluid boundary, which was later generalized by \cite{HGC+14} and \cite{BG03} by enforcing the velocity constraint not only over the boundary, but over the whole overlapping region.
In the IB method, the immersed structure is represented by a force density term in the Navier-Stokes equations. The solid and the fluid problem are solved on different, possibly non-conforming grids, whereby a Lagrangian formulation for the solid problem and an Eulerian formulation for the fluid problem is used. With respect to the methods based on moving meshes such as the Arbitrary Lagrange Eulerian method \cite{NGC+16}, no remeshing of the fluid grid is necessary and in addition, different discretization schemes may be chosen for the fluid and the solid problem (see e.g. \cite{DP12, NBZ+17_1}).

\begin{figure}
\centering
\includegraphics[width=0.8 \textwidth]{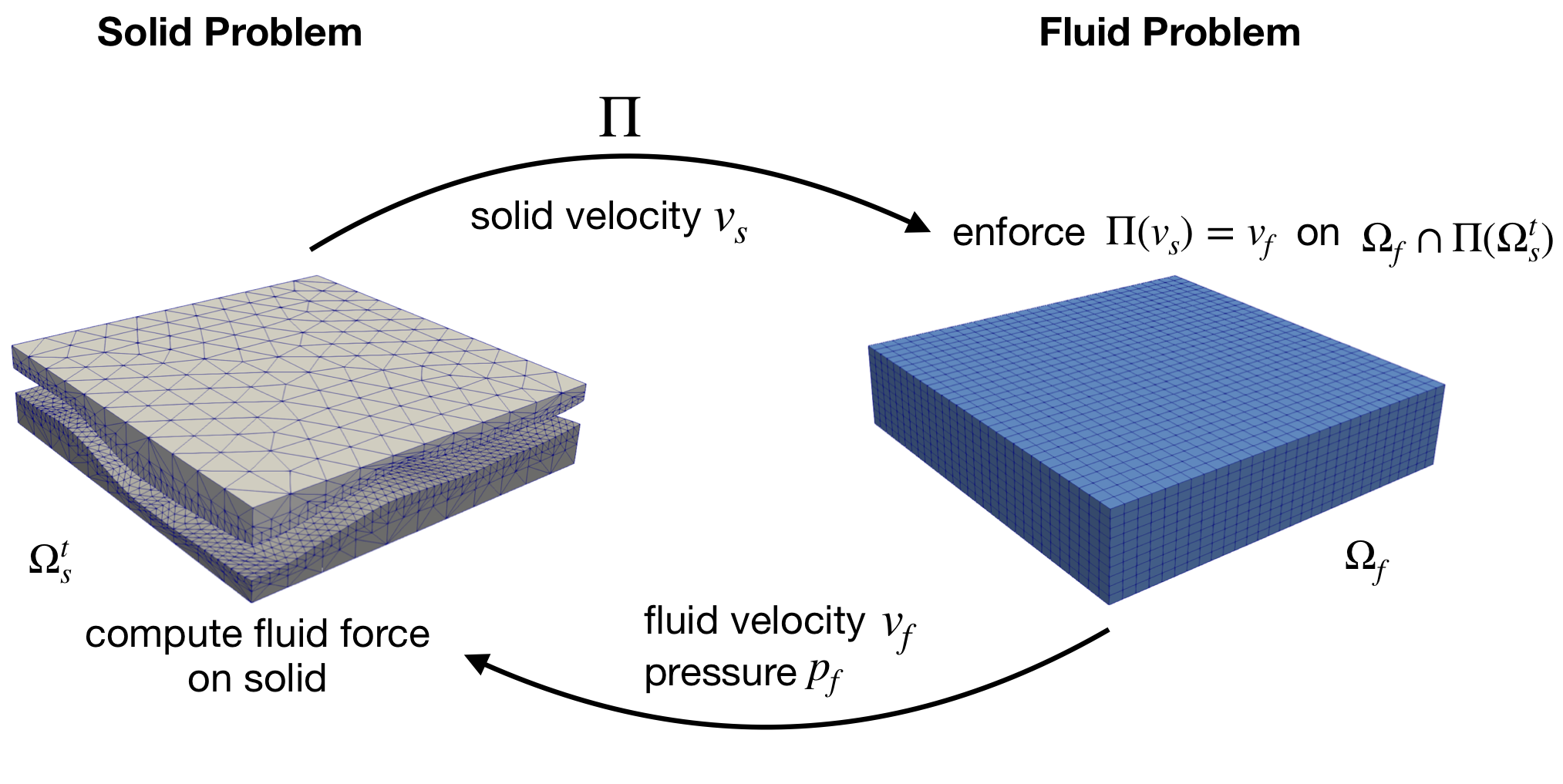} 
\label{fig:concept}
\caption{Overview of the immersed boundary approach: The solid problem is solved on the unstructured mesh on the left and the fluid problem on the structured mesh on the right. The solid domain is designed such that it can be immersed into the fluid. The two problems are coupled using variational transfer operators, here symbolized by the mapping $\Pi$.}
\end{figure}

Here, we adapt an immersed boundary approach used in cardiac simulations by \cite{NBZ+18_1}. The fluid and solid problem are solved on separate domains, whereby the solid problem is formulated in Lagrangian and the fluid problem in Eulerian coordinates.  The equality of fluid and solid velocities is enforced by a penalty method. The complete problem is solved using a staggered approach with a fixed point iteration, solving the fluid and solid problem in sequence (Fig.~\ref{fig:concept}).  For the coupling of the fluid and solid problem we use variational transfer operators, more precisely pseudo-$L^2$-projections. They are defined as projections which satisfy an orthogonality condition in the sense of the $L^2$-scalar product and ensure a stable mapping between the non-matching meshes \cite{Dic10,KZ16}.

One motivation behind the development of this type of methods was to avoid the remeshing of a moving fluid mesh in case of large solid displacements. This is necessary in other FSI approaches to ensure numerical stability of the fluid part of the simulation \cite{HGC+14}. In addition, no explicit representation of the fluid-solid boundary is required, which is a particular advantage for high resolution HM simulations. There, small scale changes in the aperture field result in large changes of the fluid flow patterns. The ability to implicitly reflect these changes in the fluid-solid boundary after every deformation, however small, immensely simplifies the setup of these simulations. Lastly, as remarked earlier, the immersed boundary approach gives some freedom which formulation, mesh type or method, can be used for solving each subproblem. These characteristics make this FSI formulation a highly suitable method to compute hydro-mechanically coupled processes in rough fractures in a coupled manner while simultaneously resolving small scale roughness on the fracture surfaces. 

This paper is organized as follows. In section \ref{sec:methods} we give an overview of the method and the governing equations. We introduce the solid and fluid problem in their strong and weak formulation, followed by the definition and construction of the transfer operators, the discretization in space and time, and finally, an explanation of the coupling and an overview of the algorithm. In section \ref{sec:results} we show numerical examples of the approach. First, we validate the method with 2D and 3D benchmarking problems and then we present numerical examples with real rough surface geometries.
\section{Method \& governing equations}
\label{sec:methods}
We use an adapted immersed boundary approach following \cite{HGC+14,NBZ+17_1,NBZ+18_1,NZK+19}. In our approach, the FSI problem is solved separately for the fluid and the solid in a fixed point iteration using finite elements. We start by solving the fluid problem formulated by the Navier-Stokes equations for incompressible flow in Eulerian coordinates. The fluid problem is coupled with the solid problem by enforcing the equality of fluid and solid velocities on the overlap between the fluid and the solid using a penalty method. We then map the fluid velocities and pressure back to the solid, convert them into a fluid stress tensor and subtract the resulting momentum from the solid balance. After solving the solid problem, formulated with linear elasticity and Lagrangian coordinates, we check for convergence of the complete iteration and either stop  or continue it by mapping the solid velocities back to the fluid (see Fig.~\ref{fig:concept} or Alg. \ref{al:algorithm}). The mappings between the domains are realized as volumetric variational transfer operators explained in subsection \ref{sec:operators}. Note, that while we employ the same principles as \cite{HGC+14}, \cite{NBZ+17_1} or \cite{NBZ+18_1}, our adaption differs from these methods, as we do not use Lagrange multipliers, but a penalty method to enforce the equality of solid and fluid velocities. 

We denote the fluid domain with  $\Omega_f \subset \mathbb{R}^3$, the moving solid domain depending on time $t$ with  $\Omega^t_s \subset \Omega_f$ , the boundary of the fluid domain with  $\Gamma_f$ and the boundary of the solid domain with $\Gamma^t_s$.  Both the fluid and the solid boundary are decomposed into a Neumann boundary with superscript N and a Dirichlet boundary with superscript D, such that we have  $\Gamma_f= \Gamma_f^D \dot{\cup} \Gamma_f^N$ and $\Gamma^t_s= (\Gamma^t_s)^D \dot{\cup} (\Gamma^t_s)^N$. The outer normal vectors on the boundaries are denoted $n_f$ and $n_s$. For function spaces we use  $L^2(\Omega)$ for the space of Lebesgue measurable and square integrable functions and $H^1(\Omega)$ for the standard space of weakly differentiable functions in $L^2(\Omega)$, where $\Omega \in \{\Omega_f, \Omega_s^t \}$.
\subsection{Strong formulation}
In this section we introduce the strong formulation of the fluid problem and the dynamic solid problem. The fluid problem is formulated using the Navier Stokes equations for incompressible flow and the solid problem with linear elasticity.

We denote with $\sigma_f$ the fluid stress tensor, with $\rho_f$ the fluid density, with $v_f$ the fluid velocity and with $\mu_f$ the dynamic fluid viscosity. For the solid problem we introduce the stress tensor $\sigma_s$ of the solid, the solid displacement $u_s$ and the solid density $\rho_{s}$. For time derivatives, we use the dot notation, i.e. $\dot{u}_s := \frac{\partial}{\partial t}u_s$.  The strong formulation of the FSI problem then reads as:
\begin{align}
\rho_f  \dot{v}_f  + \rho_f ( v_f \cdot \nabla) v_f  - \mu_f \nabla \cdot \sigma_f(p_f, v_f) &= f_{\text{fsi}} &\qquad  \text{ on } \Omega_f,  \label{eq:navier} \\
\nabla \cdot v_f &= 0 & \qquad \text{ on } \Omega_f, \label{eq:stokes} \\
v_f &= v_{f,0} &\qquad \text{ on }  \Gamma_f^D, \label{eq:bc_f_dc} \\ 
\sigma_f \cdot n_f &= h &\qquad \text{ on }  \Gamma_f^N, \label{eq:bc_f_nm} \\
\rho_{s}  \ddot{ u}_s  - \text{div}\sigma_s(u_s) &= f_s &\qquad \text{ on } \Omega^t_s, \label{eq:eldynamics} \\ 
u_s &= 0 &\qquad \text{ on } (\Gamma^t_s)^D, \label{eq:solid_dc} \\ 
\sigma_s \cdot n_s &= t &\qquad \text{ on } (\Gamma^t_s)^N, \label{eq:solid_nm} \\
\dot{u}_s &= v_f &\qquad \text{ on } \Omega^t_s \cap \Omega_f. \label{eq:strongcoupling} 
\end{align}
Equations \ref{eq:navier} and \ref{eq:stokes} are the Navier Stokes equations with  Dirichlet and Neumann (Eq. \ref{eq:bc_f_nm}) boundary conditions (Eq.\ref{eq:bc_f_dc}). Equation \ref{eq:eldynamics} is the equation of motion for the solid, which consists of a linear elastic material obeying Hooke's law. Again we have Dirichlet (Eq. \ref{eq:solid_dc}) and Neumann (Eq. \ref{eq:solid_nm}) boundary conditions. Equation \ref{eq:strongcoupling} couples the fluid and solid velocities on the overlap $\Omega^t_s \cap \Omega_f$. The relationship between the Cauchy stress $\sigma_f$ for a Newtonian fluid and fluid velocity and pressure is given by:
\begin{equation}
\sigma_f = -p_f \text{Id} + \frac{{\nabla v_f}^t + \nabla v_f} {2 } \label{eq:cauchy_stress_fluid}.
\end{equation}

In the following, we will interchangeably use $v_s  :=  \dot{u}_s = \frac{\partial}{\partial t}u_s $ for the solid velocity. For the sake of preserving the generality and extendability of our approach, we use time dependent formulations here. However, it is worth mentioning, that the problem could also be formulated in steady state.

\subsection{Weak Formulation}
We describe the weak formulation for two reasons in particular: First, in finite elements it is ultimately an approximation of the weak formulation that is implemented and secondly, it is by the weak formulation that the second part of the coupling described later in subsection \ref{sec:coupling} can be understood.

We begin by introducing the function spaces for the velocities $v$ and pressure $p$ for the fluid problem:
\begin{align}
\mathcal{W}^v &= \big \{ \delta v \in [H^1(\Omega_f)]^d \, | \, \delta v = 0 \text{ on } \Gamma_f^D \big \}, \; d=2,3 \\
\mathcal{W}^p &=  [L^2(\Omega_f)]^d,  \; d=2,3.
\end{align}
For test functions $\delta v \in \mathcal{W}^v$ and $\delta p \in \mathcal{W}^p$ we can then write the fluid problem in the following weak form:
\begin{eqnarray}
\int_{\Omega_f} \rho_f \,  \big (\dot{v}_f +  (v_f \cdot \nabla  ) \, v_f \big ) \, \cdot \delta v \, dV + 
\int_{\Omega_f}    \sigma_f  : \nabla ( \delta v) \, dV & -  \nonumber \\
\int_{\Omega_f}   \rho_f f_{\text{fsi}} \cdot \delta v \, dV  - \int_{\Gamma_f^N}  h \cdot \delta v \, dA +
\int_{\Omega_f}  \nabla v \cdot \delta p \, dV &= 0. \label{eq:fluid_weak}
\end{eqnarray}
For the solid problem we use the following space for the displacements:
\begin{equation}
\mathcal{V}^{u} = \big \{ \delta u \in [H^1(\Omega_s^t)]^d \, | \, \delta u = 0 \text{ on } (\Gamma_s^t)^D \big \}, \; d=2,3
\end{equation}
and write it in the weak form:
\begin{equation}
\rho_s  \int_{\Omega^t_s}  \ddot{u}_s \cdot \delta  u \, dV +
\int_{\Omega^t_s} \sigma_s : \nabla(\delta u)  \, dV -
\rho_s   \int_{\Omega^t_s} \, f_s \cdot \delta u  \, dV - \int_{(\Gamma^t_s)^N}  t \cdot \delta u \, dA =0 . \label{eq:solid_weak}
\end{equation}
\subsection{Discretization in space and time}
\label{sec:discretization}
For both the fluid and the solid problem we use a Galerkin finite element method. We mesh $\Omega_f$ with a structured rectangular mesh of hexahedral elements and denote the corresponding space of bilinear Lagrange elements for the fluid velocities by $W^{v}$ and the space of bilinear Lagrange elements  for the fluid pressure with $W^{p}$. Furthermore, we employed Streamline Upwind Petrov Galerkin (SUPG) and the Pressure Stabilized Petrov Galerkin (PSPG) formulations for stabilization. The solid domain $\Omega_s^t$ has an unstructured mesh with tetrahedral elements and the corresponding  spaces of linear Lagrange elements  for the solid pressure, displacements and velocities are denoted $V^{p}$,  $V^{u}$ and $V^{v}$. The degrees of freedom for each space are denoted with $n^{\bullet},\, \bullet \in \{W^{v}, W^{p}, V^{u}, V^{p}, V^{v}\}$.
Both the solid and the fluid problem are time dependent.  For the fluid problem we used backward differentiation formula, BDF2 and for the solid problem we used a Newmark scheme with parameters $\beta=\frac{1}{2}$ and $\gamma = \frac{1}{4}$.
\subsection{Variational Transfer Operators}
\label{sec:operators}
One of the main characteristic of our method is that we solve the fluid and the solid problem on different domains with non-matching meshes. The two problems need to be coupled however, i.e. we have to transfer velocities and pressures between the two domains. This is achieved by using variational transfer operators, which provide a stable and flexible way to map variables between arbitrary meshes.

To define the operators, we assume that the solid domain $\Omega_s^t$ is immersed into the fluid domain $\Omega_f$. We further assume that we have a function space $V$ on $\Omega_s^t$ with elements $v$ and a function space $W$ on $\Omega_f$ with elements w, such that $V$ is a subset of $W$, i.e. $V \subset W$ and that the $L^2$-scalar product $(v,w)_{L^2} = \int_{\Omega_f} v\, w \, d \omega$ is well defined. The variational transfer operator is then defined as a mapping $\Pi : V \rightarrow W$, such that for every $v \in V$ the following $L^2$-orthogonality holds for all elements $\mu$ of a yet to be defined multiplier space $M$:
\begin{align}
\int\limits_{\Omega_f}  \big(\Pi(v) - v \big) \mu \, d \omega = 0 \quad \forall \mu \in M. \label{eq:l2-proj}
\end{align}
Naturally, in applications we need a discrete matrix representation $T$ of $\Pi$.  $T$ is constructed by reformulating Eq.\ref{eq:l2-proj} and introducing the discrete approximations of the spaces $V, W$ and $M$. These discrete spaces are $V_h, W_h$ and $M_h$ with bases $(\phi_i^{V_h})_{i=1,...,n^{V_h}}$,  $(\phi_i^{W_h})_{i=1,...,n^{W_h}}$ and  $(\phi_i^{M_h})_{i=1,...,n^{M_h}}$,  where $n^{V_h}$, and $n^{W_h} = n^{M_h}$ are the dimensions of each space. Now, let $v_h$ be an element of $V_h$, then with $w_h:= \Pi(v_h)$, we can reformulate Eq. \ref{eq:l2-proj} as:
\begin{align}
\int\limits_{\Omega_f} w_h \, \mu_h \, d \omega = \int\limits_{\Omega_f} v_h \, \mu_h \, d \omega,  \quad \forall \mu_h \in M_h, 
\end{align}
and using the basis representations of $v_h, w_h$ and $\mu_h$ we get:
\begin{align}
\int\limits_{\Omega_f} \sum_{i=1}^{n_{W_h}} w_i \phi_i^{W_h} \, \phi_k^{M_h} \, d \omega = \int\limits_{\Omega_f} \sum_{j=1}^{n_{V_h}} v_j \phi_j^{V_h} \, \phi_k^{M_h} \, d \omega,  \quad k = 1, ...., n_{M_h} \\
\Leftrightarrow  \sum_{i=1}^{n_{W_h}}  w_i \int\limits_{\Omega_f} \phi_i^{W_h} \, \phi_k^{M_h} \, d \omega = \sum_{j=1}^{n_{V_h}} v_j \int\limits_{\Omega_f} \phi_j^{V_h}  \, \phi_k^{M_h} \, d \omega,  \quad k = 1, ...., n_{M_h}. 
\end{align}
With
\begin{align}
\mathbf{w} & := (w_i)_{i=1,..., n^{W_h}}  \quad \text{and} \quad \mathbf{v} := (v_j)_{j=1,..., n^{V_h}},  \\
D &:= (d_{ik})_{i,k=1,...,n^{W_h}} , \quad   d_{ik} := \int\limits_{\Omega_f}  \phi_i^{W_h} \, \phi_k^{M_h} \, d \omega \, , \label{eq:op_ass_d}\\
B &:=(b_{jk})_{j=1,..., n^{V_h}, k=1,..., n^{M_h}}, \quad   b_{jk} := \int\limits_{\Omega_f}  \phi_i^{V_h} \, \phi_k^{M_h} \, d \omega , \label{eq:op_ass_b}
\end{align}
we can now define the discrete operator as $T:= D^{-1} B$ and obtain the relationships: 
\begin{equation}
D \mathbf{w} = B \mathbf{v} , \, \text{or} \,  \mathbf{w} = D^{-1} B \mathbf{v} = T \mathbf{v} .
\end{equation}
Note that $D$ and $B$ are essentially volumetric mass matrices, whereby $D$ is a quadratic and $B$ a rectangular matrix . We have some liberty, which multiplier space  $M_h$ we select. With the most straight forward choice $M_h = W_h$ we would end up computing the inverse of $D$, which is an unwanted computational expense. Instead we choose a space $M_h$ spawned by biorthogonal basis functions \cite{Woh00}, which leads to a diagonal matrix $D$ that is easy to invert. More specifically, in this use, operators of this type are called pseudo-$L^2$-projections. They have a solid mathematical base and $H^1$-stability and $L^2$-approximation property are shown in \cite{Dic10}. Following this principle, we assemble the projections:
\begin{align}
\mathbf{T}_{v_s} &:V^{v} \rightarrow W^{v}\; \text{, for transferring solid velocities to the fluid domain}, \\
\mathbf{T}_I &:V^{i} \rightarrow W^{i}\; \text{, for transferring an indicator variable to the fluid domain}, \\
\mathbf{T}^*_{p_f}&:W^{p} \rightarrow V^{p}\; \text{, for transferring fluid pressure to the solid domain}, \\ 
\mathbf{T}^*_{v_f} &:W^{v} \rightarrow V^{v}\; \text{, for transferring fluid velocities to the solid domain}.
\end{align}
In practice, the assembly of the discrete variational operators is highly complex. To compute the individual entries $d_{ik}$ and  $b_{ik}$ in Eq. \ref{eq:op_ass_d} \& \ref{eq:op_ass_b}, we need to know all intersections between the elements of the fluid and the solid mesh, in order to determine the common support of the bases $\phi^{W_h}, \phi^{V_h}$ and $\phi^{M_h}$. In parallel computing environments these elements might lie on different compute nodes, so additional communication between the nodes is necessary.  It is only with the recent development of the MOONoLith library \cite{KZ16}, that the assembly routines have become efficient enough to make this immersed boundary approach practicable.
\subsection{Coupling}
\label{sec:coupling}
The fluid and the solid problem are fully coupled. First,  we enforce the equality of fluid and solid velocities on the overlap of the immersed solid and the fluid using a penalty method. Secondly, we transfer the fluid velocity and pressure to the solid problem and use them to compute a fluid force term. Note that the coupling relies on the transfer operators which are assembled by computing volumetric integrals. Hence no explicit representation of the boundary between the fluid and the solid is ever required.

To couple the fluid with the solid we apply a penalty method, which enforces  the equality of  fluid and solid velocities. Let $\epsilon$ be then penalty parameter and $\Pi_{\dot{u}_s}$the non-discrete form of $\mathbf{T}_{\dot{u}_s}$. Then we add the following term to the weak formulation of the fluid problem (Eq. \ref{eq:fluid_weak}):

\begin{equation}
\int_{\Omega_f \cap \Omega_s} \epsilon^{-1}\cdot \left(\Pi_{\dot{u}_s} (\dot{u}_s) - v_f \right) \delta v \, dw. \label{eq:penalty}
\end{equation}

In its discrete form, the overlap between the solid and the fluid is determined by constructing the transfer operator $\mathbf{T}_I: V^i \rightarrow W^i$, which maps an indicator variable $i$ from the solid to the fluid mesh, whereby  $V^i$ and $W^i$ are the spaces of (bi-)linear Lagrange elements on each domain. Let $\mathcal{T}^W$ be the mesh of the space $W^i$. The overlap $\Omega_f \cap \Omega_s$ is defined as all nodes $p$ in $\mathcal{T}^{W}$ with $T_I(i) \geq \alpha$  ($\alpha$ being a mesh dependent threshold value). 

To couple the solid with the fluid, we transfer the fluid velocity $v_f$ and pressure $p_f$ to the solid simulation after every iteration. We then recompute the Cauchy stress according to Eq. \ref{eq:cauchy_stress_fluid} and subtract the term for the contributions of the fluid forces, i.e. the second term $\int_{\Omega_f} \sigma_f  : \nabla ( \delta v) \, dV$
in Eq. \ref{eq:fluid_weak}, from the weak formulation of the solid (Eq. \ref{eq:solid_weak}). In addition, we should also subtract the dynamic contributions of the fluid phase from the solid phase.  But since in geophysical applications the Young's Modulus of the solid is normally on the order of GPa and solid velocities are consequentially small,  we omit the dynamic contributions of the fluid on the solid.

Note that using a penalty method can be  seen as a simplification of the coupling described in \cite{HGC+14} and \cite{NBZ+18_1}. There, Lagrange multipliers are used, which can be interpreted as forces acting on the fluid. From a theoretical point of view, using Lagrange multipliers is superior to penalty methods as the forces can be mapped back from the fluid to the solid using the transposed of the mapping operators, resulting in a variationally consistent scheme. There are, up to the knowledge of the authors, no equivalent results  for using a penalty method. On the other hand,  the penalty method has  the advantage that no saddle point problem needs to be solved, instead we can use a standard iterative solver for the regularized problem. Furthermore, we will show in the section \ref{sec:results} with numerical experiments, that using the penalty method does not impact the capability of resolving the solid-fluid boundary.
\subsection{Algorithm}
\label{sec:algorithm}
The coupled problem is solved in a fixed point iteration in up to $L_{\mathrm{FP}}$ steps. We formulate one step $l$ by introducing the coupling terms to the weak formulations of Eq. \ref{eq:fluid_weak} - \ref{eq:solid_weak}. For the solid problem we define the operator $S(u):= \mathcal{V}^{u} \rightarrow (\mathcal{V}^{u})^*$ as:

\begin{equation}
    (S(u), \delta u):= \rho_s  \int_{\Omega^t_s}  \ddot{u}_s \cdot \delta u \, dV + \int_{\Omega^t_s} \sigma_s : \nabla(\delta u)  \, dV, \label{eq:S}
\end{equation}
and the right hand side $R_s(\delta u)$ as:
\begin{equation}
R_s(\delta u) = \rho_s   \int_{\Omega^t_s}  f_s \cdot \delta u  \, dV - \int_{(\Gamma^t_s)^N}   t  \cdot \delta  u \, dA + \int_{\Omega_f \cap \Omega_s}    \sigma_f  : \nabla ( \delta u) \, dV. \label{eq:rhs_s}
\end{equation}
Note, that we have added the contributions from the fluid $\int_{\Omega_f \cap \Omega_s}    \sigma_f  : \nabla ( \delta v) \, dV$ to $R_s$. This will become relevant  later in the context of a fixed point iteration, as we will use the results of a previous iteration $l-1$ to form $\sigma_f$.
For the fluid problem we define $ F(v,p):= \mathcal{W}^v \times \mathcal{W}^p  \rightarrow (\mathcal{W}^v \times \mathcal{W}^p)^*$ as:
\begin{eqnarray}
(F(v,p), (\delta v, \delta p)) = \rho_f \int_{\Omega_f} \,  \big (\dot{v}_f +  (v_f \cdot \nabla  ) \, v_f \big) \, \delta v \, dW + 
\int_{\Omega_f}    \sigma_f  : \nabla ( \delta v) \, dW & +  \nonumber \\
\int_{\Omega_f}  \delta v \nabla v \, dW \label{eq:F}
\end{eqnarray}
and the right hand side $R_f(\delta v)$ as:
\begin{eqnarray}
R_f(\delta v) = \int_{\Gamma_f^N}  h  \cdot \delta v\, dA  +  \rho_f \int_{\Omega_f}   f_{\text{fsi}} \cdot \delta v  \, dW. 
\label{eq:rhs_f}
\end{eqnarray}
Using the discretizations from Section \ref{sec:discretization} we obtain the discrete representatives $\mathbf{S}, \mathbf{F}, \mathbf{R_s}$ and $\mathbf{R}_f$ from $S$(Eq.\ref{eq:S} ), $F$(Eq.\ref{eq:F}), $R_s$(\ref{eq:rhs_f}), and $R_f$(Eq.\ref{eq:rhs_f}), respectively. We then discretize the penalty term from Eq. \ref{eq:penalty} and split it into two parts $\mathbf{P}_s$ and $\mathbf{P}_f$.  Then we denote with $\mathbf{y}_s$ and $\mathbf{y}_f$ the solutions of the solid and fluid problem, whereby $\mathbf{y_f}$ contains fluid velocities and pressure, and write the coupled problem as:

\begin{equation}
\left [
\begin{array}{cc} 
\mathbf{S} & \mathbf{0} \\ 
\mathbf{P}_s & \mathbf{F} - \mathbf{P}_f
\end{array}
\right] 
\left[
\begin{array}{c}
\mathbf{y}_s \\
\mathbf{y}_f
\end{array}
\right]
=
\left[
\begin{array}{c}
\mathbf{R}_s \\
\mathbf{R}_f
\end{array}
\right],
\end{equation}
where $\mathbf{P}_s$is defined as:
\begin{equation}
\mathbf{P}_s := \epsilon^{-1} 
\left [
\begin{array}{c} 
\mathbf{I}_{\Omega_f \cap \Omega_s} \cdot \mathbf{M}_f \cdot \mathbf{T}_{v_s}  \\
\mathbf{0}
\end{array}
\right],
\end{equation}
and $\mathbf{P}_f$ as:
\begin{equation}
\mathbf{P}_f := \epsilon^{-1} \left[
\begin{array}{cc}
\mathbf{I}_{\Omega_f \cap \Omega_s} \cdot \mathbf{M}_f & \mathbf{0}\\
\mathbf{0} & \mathbf{0}
\end{array}
\right].
\end{equation}
Here, $\mathbf{I}_{\Omega_f \cap \Omega_s}
$ is a diagonal matrix containing zeros and ones according to the values of $T_I(i)$, i.e.:
\begin{equation}
(\mathbf{I}_{\Omega_f \cap \Omega_s})_{k_l} := \left\{
                \begin{array}{ll}
                1, \; \text{if}\, k=l, \text{and}\, T_I(i) > t, \\
                0, \, \text{otherwise}
                \end{array}
                \right.
\end{equation}
where $l,k=1,...,n^{W^v}$ and $\mathbf{M}_f$ is the mass matrix obtained from the discretization on $W^v$. The matrices $\mathbf{0}$ contain all the entries associated with the degrees of freedom from the fluid pressure. By adding the superscripts $l, l-1$ for the iteration steps of the fixed-point iteration, we can formulate the staggered approach to solve the coupled problem by algorithm \ref{al:algorithm}.


\begin{algorithm}
\caption{Fixed point iteration}

\begin{algorithmic}[1]
\STATE Initialization: $l=1, u_s=u_{s,0}, v_f=v_{f,0}, p_f=p_{f,0}$
\FOR {$l=1,2,...,L_{FP}$}
	\STATE Transfer $v_f$ and $p_f$ to the solid domain by setting $\tilde{v}_f := \mathbf{T}_{v_f}^* v_f$ and $\tilde{p}_f := \mathbf{T}_{p_f}^* p_f$.
  \STATE Form $\mathbf{R}_s^{l-1}$ by assembling $\sigma_f$, using $\tilde{v}_f$ and $\tilde{p}_f$ (cf. Eq. \ref{eq:rhs_s}).
  \STATE Solve solid problem: $\mathbf{S} \cdot \mathbf{y_s}^l = \mathbf{R}_s^{l-1}$.
  \IF{$\frac{\|\mathbf{y}_s^{l-1} - \mathbf{y}_s^l\|}{\|\mathbf{y}_s^{l-1}\|} \leq \mbox{tol}_{\mathrm{FP}}$}
  \STATE STOP.
  \ENDIF
  \STATE Solve fluid problem: $(\mathbf{F} - \mathbf{P}_f) \cdot \mathbf{y_f}^l  = \mathbf{R}_f^{l-1} - \mathbf{P}_s \cdot \mathbf{y_s}^l$.
\ENDFOR
\end{algorithmic}
\label{al:algorithm}
\end{algorithm}

We note that the fixed-point iteration is carried out in each time-step. For ease of presentation, here we have dropped any indices connected to the time discretization. We furthermore note that the transfer of the solid velocities is realized by means  of the multiplication of $\mathbf{y}_s$ with $\mathbf{P}_s$, as the latter  contains the discrete projection $\mathbf{T}_{v_s}$. The fixed-point iteration shown in Alg. \ref{al:algorithm} can be interpreted as a block-Gauss-Seidel iteration. We will see later in sec. \ref{sec:benchmarks} that in practice convergence of the overall FSI problem is only mildly dependent on $\epsilon$.

\subsection{Implementation}
The variational transfer operators are implemented as userobjects in the MOOSE framework \cite{GNH+09}, employing MOONoLith \cite{KZ16}, Utopia \cite{ZKN+16}, and libMesh \cite{KPH+06}. The MOOSE framework is also used to form the weak formulations. All components are designed for parallel computing.
\section{Numerical examples}
\label{sec:results}
The following section validates the presented numerical modeling approach with 2D parallel plate (Section~\ref{sec:bm_parallel}) and 3D tunnel flow (Section~\ref{sec:bm_tunnel}) benchmark problems, before applying the approach to real fracture geometries (Section~\ref{sec:roughFractureSurfaces}).

Throughout the simulations, we use the same material parameters, which are representative of a hard granodiorite rock, saturated with water.  For the rock we use a Poisson ratio of 0.33, a Young's modulus of 1\e{10}~Pa, and a density of 2.75~$g/cm^3$. For water at about 20$^\circ$C we use a kinematic viscosity of 1~${mm}^2/s$ and a density of 1~$g/cm^3$. The penalty parameter $\epsilon$ was set to $10^{-6}$, the tolerance $tol_{\mathrm{FP}}$ to $10^{-8}$ and $L_{\mathrm{FP}}$ to 10.
Simulations were conducted on the cluster of the Institute of Computational Science in Lugano, Switzerland, using up to 10 compute nodes (2~x Intel Xeon E5-2650~v3 @~2.30GHz) with 16 CPU's each.

\subsection{Benchmark problems}
\label{sec:benchmarks}
We designed a two benchmark simulations to assess the method's capability of detecting boundaries and simulating realistic fluid flow.  The first benchmark is a 2D parallel plates example and a second one a simulation of tunnel or Poiseuille flow in 3D.  The results of the simulations are compared to analytical solutions and/or the solution of an equivalent setup for fluid flow with rigid boundaries.

\begin{figure}[hbt!] 
\centering             
\includegraphics[width=0.8\textwidth]{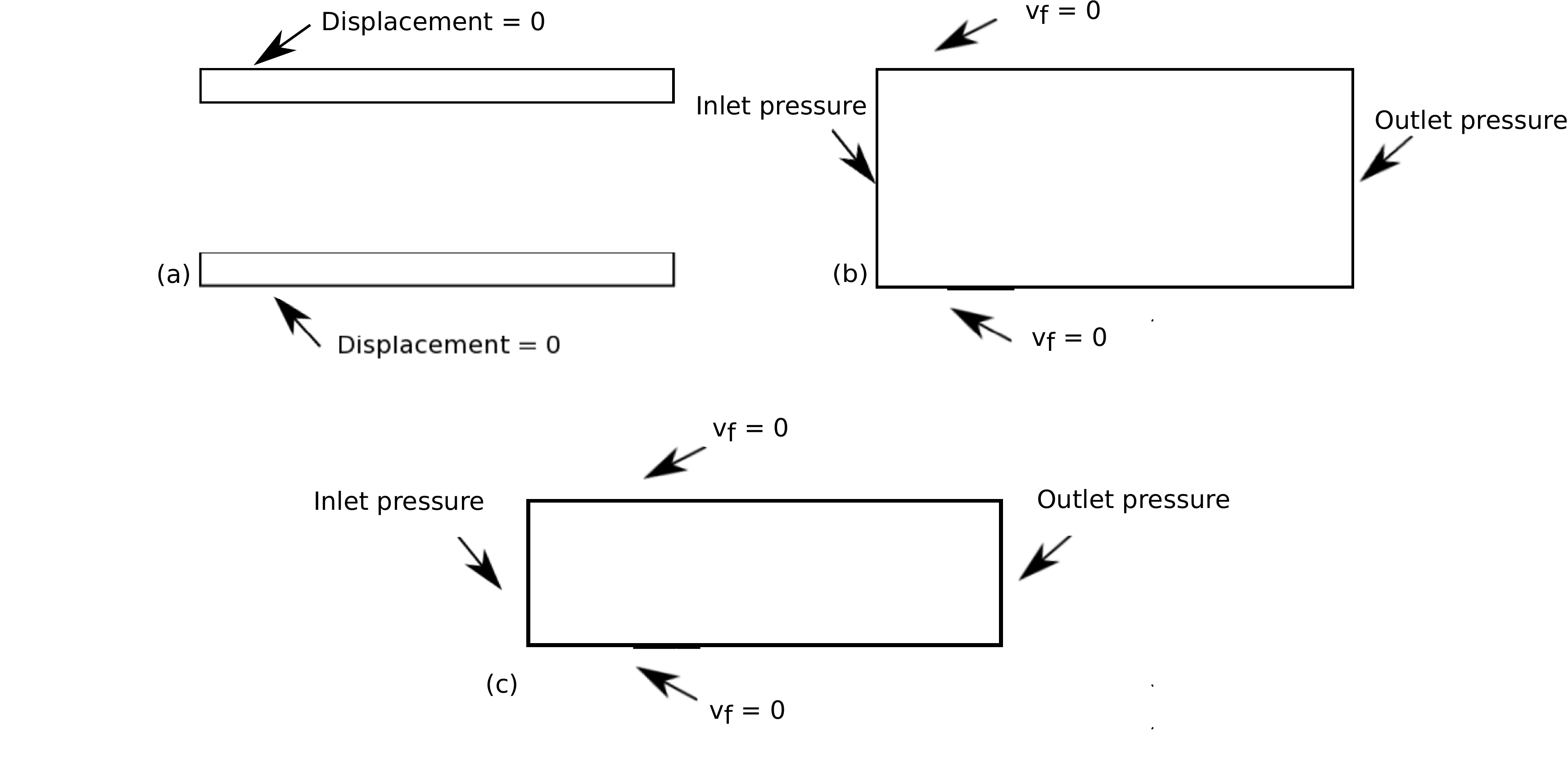}
\caption{Boundary conditions of the 2D parallel plate simulation: a) Solid mesh of the FSI problem; b) Fluid mesh of the FSI problem; and c) Equivalent Navier-Stokes simulation.}
\label{fig:bnmrk_setup}
\end{figure}

Apart from the dimension and the used geometries, the equivalent Navier-Stokes simulation experiments have the same setup.  We apply zero-Dirichlet boundary conditions on the entire solid boundary, thereby creating a rigid object.  On the fluid domain we apply a pressure gradient of 0.1~MPa from the left inlet to the right outlet, forcing the fluid to flow (Fig.~\ref{fig:bnmrk_setup}).

\begin{figure}[hbt!] 
\centering             
\includegraphics[width=0.8\textwidth]{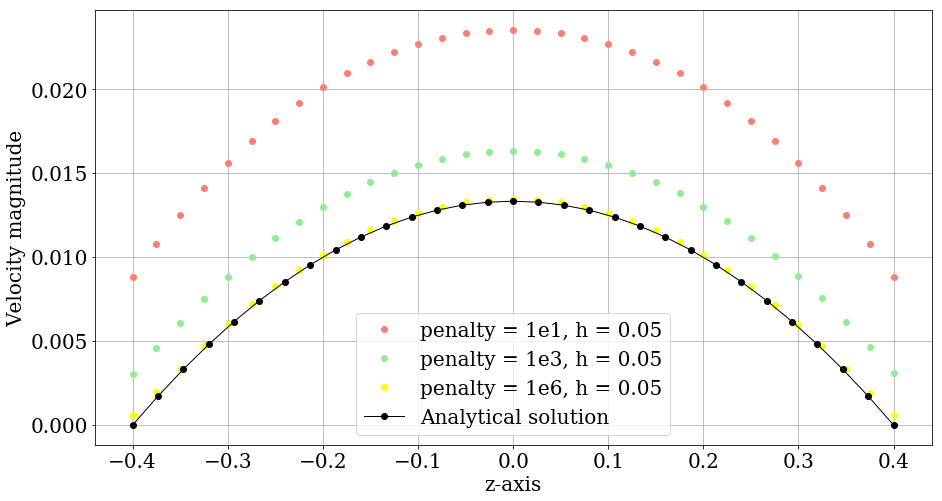}
\caption{Solutions for different choices of penalty  parameter $\epsilon$.}
\label{fig:bnmk_epsilon}
\end{figure}

\begin{figure}[hbt!] 
\centering          
\includegraphics[width=0.6\textwidth]{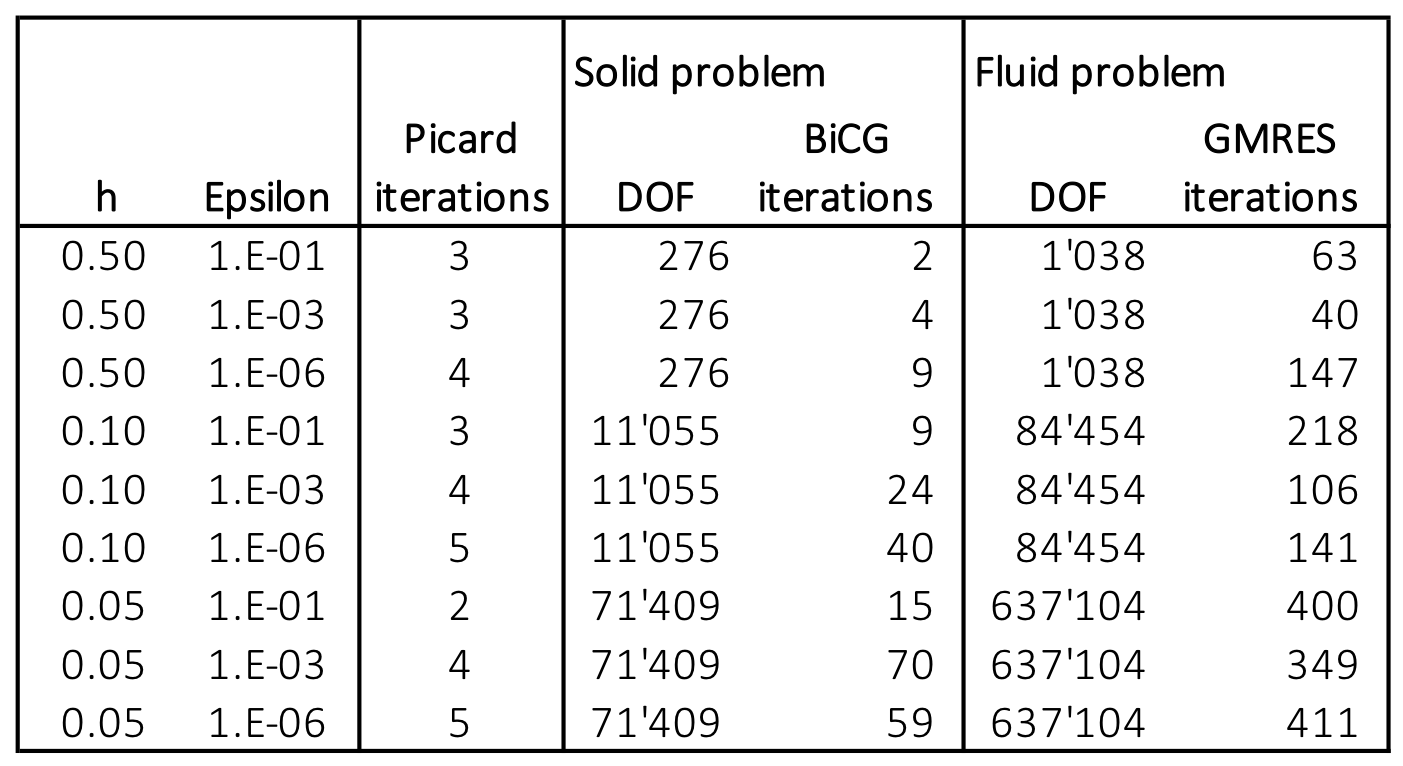}
\caption{Iteration count for different combinations of $h$ and $\epsilon$ (DOF: degrees of freedom).}
\label{fig:bnmk_iterations}
\end{figure}

The benchmark problems were also used to investigate the influence  of the penalty parameter. We refer to  Fig.~\ref{fig:bnmk_epsilon}, where we ran the 3D simulation for tunnel flow with the different penalty parameters $\epsilon=10^{-1}, 10^{-3}, 10^{-6}$, and compared the resulting discrete solutions  to the analytical solution. Eventually, the penalty parameter $\epsilon$ was set to $10^{-6}$.
In addition to solving the FSI problem with different penalty parameters, we also ran the 3D tunnel flow for different mesh widths $h$ (with $h=h_f \approx h_s$) to assess how the convergence of the overall and the fluid and the solid problem depended on $h$ and $\epsilon$. The results  collected in Fig.~\ref{fig:bnmk_iterations} show only a mild dependency on $\epsilon$ in terms of the number of necessary fixed-point iterations. For the inner linear problems we used Krylov-space methods, more precisely a biconjugate gradient method (BiCG) for the solid problem and GMRES for the fluid problem. We note that by using standard preconditioners for the solid problem, such as, e.g., algebraic multigrid methods, the increase in the iteration numbers could be avoided. 


\subsubsection{2D Parallel plate} \label{sec:bm_parallel}
\begin{figure}[hbt!] 
\centering             
\includegraphics[width=0.8\textwidth]{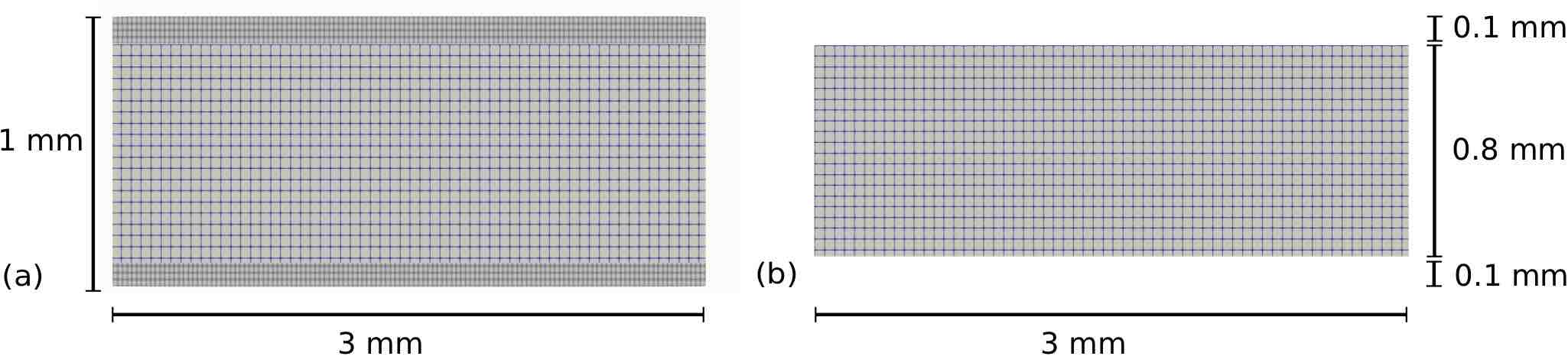}
\caption{2D parallel-plate meshes: a) Combined solid and fluid mesh of the immersed boundary approach; and b) Fluid mesh of the equivalent Navier-Stokes setup.}   
\label{fig:bnmrk_2d_setup}
\end{figure}

\begin{figure}[hbt!] 
\centering             
\includegraphics[width=0.8\textwidth]{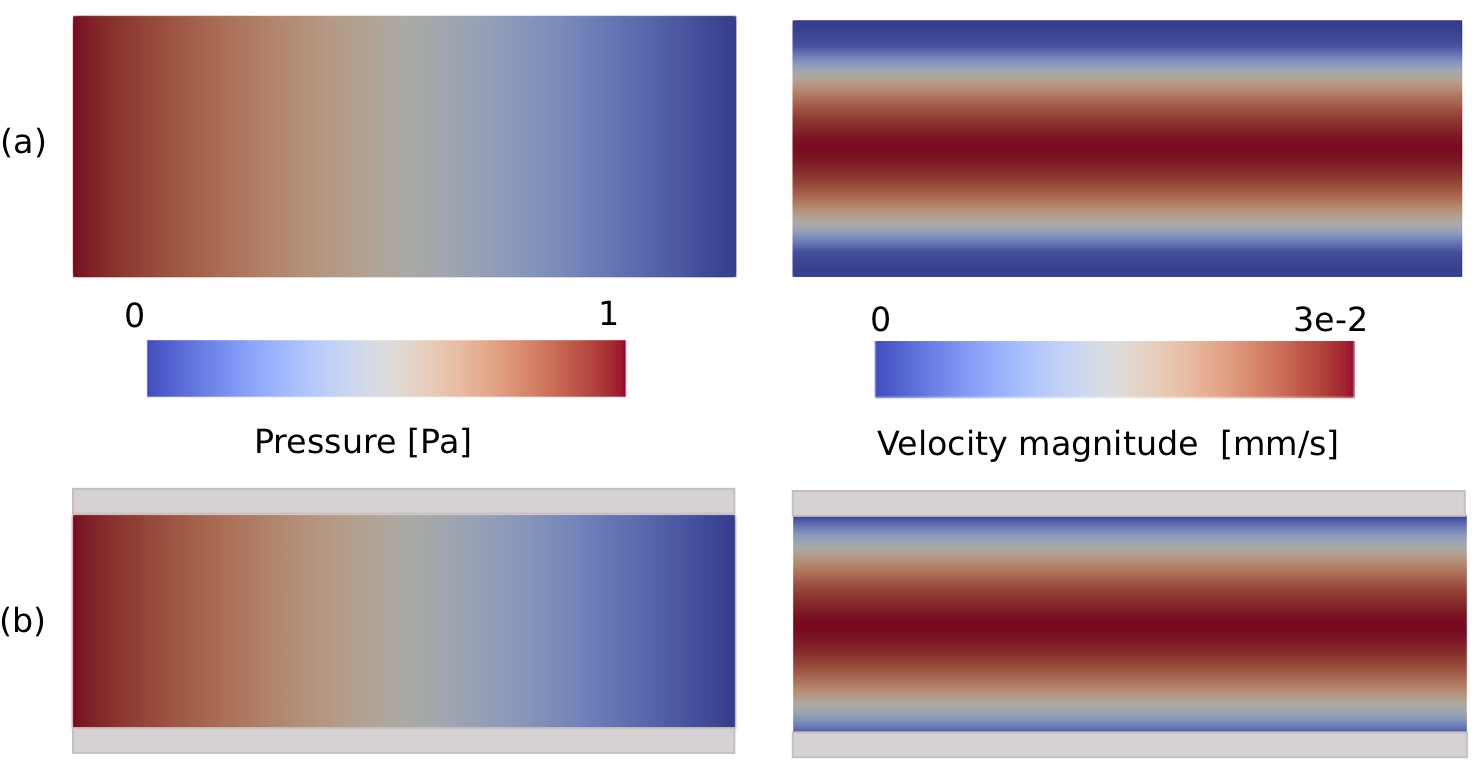}
\caption{Fluid pressure (left column) and fluid velocity (right column) fields for the 2D parallel plate simulation: a) Immersed boundary setup; and b) Equivalent Navier-Stokes setup.}
\label{fig:2d_cmp_vel}
\end{figure}

\begin{figure}[hbt!] 
\centering             
\includegraphics[width=0.8\textwidth]{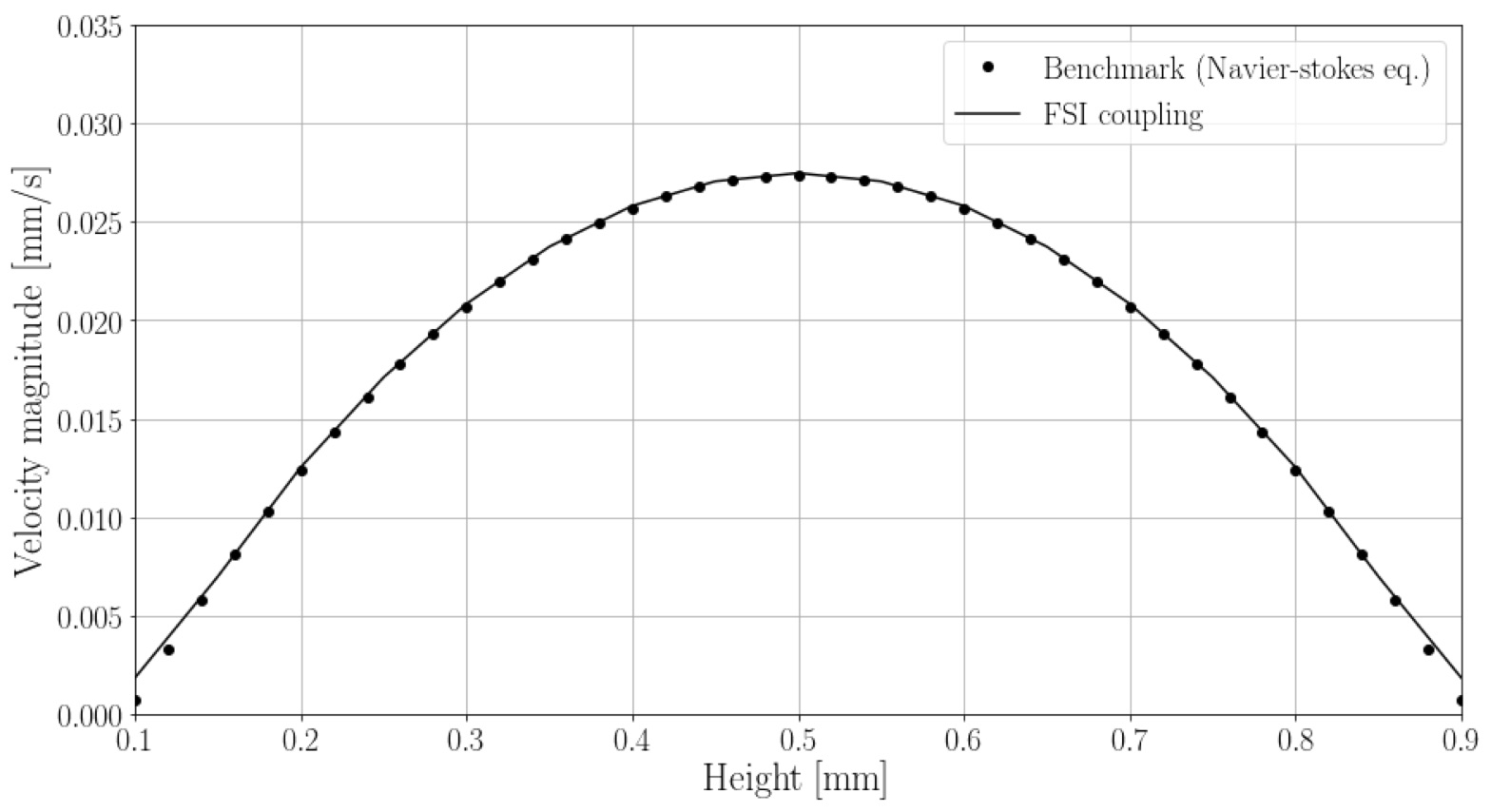}
\caption{Velocity profiles at $x = 1.5$ mm of the 2D parallel plate simulation.}
\label{fig:2d_cmp_anls_exp}
\end{figure}

In this experiment, the fluid domain has a length of 3~mm and a width of 1~mm.  The immersed solid domain consists of two parallel plates with a thickness of 0.1~mm each, located at the boundary of the fluid domain, forming a channel (Fig.~\ref{fig:bnmrk_2d_setup}a).  The solid mesh consists of 1'270 nodes with mesh width $h_s$ of 0.025~mm. The fluid mesh has 1'281 nodes and a mesh width $h_f$ of 0.08~mm. As a comparison and proxy for an analytical result, we set up an equivalent Navier-Stokes finite element simulation. Its length is 3~mm and in order to mimic the same solid boundary as in our benchmark experiment, the width is 0.8~mm. We apply the same pressure gradient and set the fluid velocity to zero across the lateral boundaries (Fig.~\ref{fig:bnmrk_2d_setup}b). The resulting fluid velocities in the center of the channel (i.e. at x=1.5~mm) are compared between the immersed boundary approach and the classic setup and show good agreement (Fig.~\ref{fig:2d_cmp_anls_exp}). See also Fig.~\ref{fig:2d_cmp_vel} for global comparison.


\subsubsection{3D Tunnel-flow} \label{sec:bm_tunnel}
To test the immersed boundary approach in three dimensions, we designed a benchmark problem of 3D tunnel flow.  As this is a classic set-up of Poiseuille flow, we can compare the results of the immersed boundary approach with the analytical solution.

\begin{figure}[hbt!] 
\centering             
\includegraphics[width=0.8\textwidth]{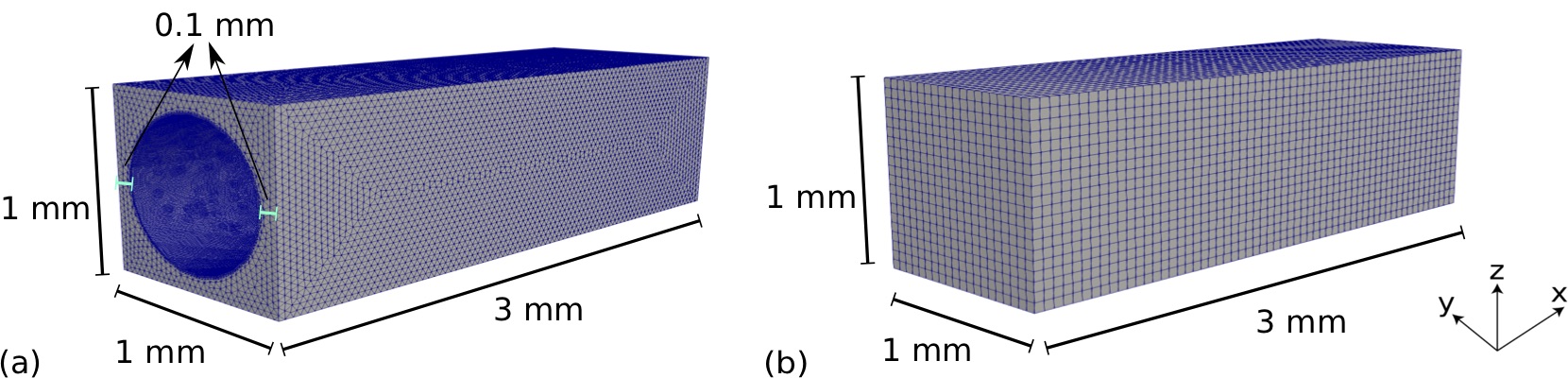}
\caption{Geometries for 3D channel flow for the immersed boundary approach: a) Solid mesh; and b) Fluid mesh.}
\label{fig:setup_3d_bnmrk}
\end{figure}

The fluid domain has a length of 3~mm and a width of 1~mm (Fig.~\ref{fig:setup_3d_bnmrk}b).  The structured mesh has 203'401 nodes with a mesh width $h_f$ of 0.05~mm.  The solid domain has the same length and width as the fluid domain (Fig.~\ref{fig:setup_3d_bnmrk}a), but along the x-axis, we have cut out a cylinder with a diameter of 0.4~mm. The unstructured mesh has 63'866 nodes and a mesh width $h_s$ of around 0.02~mm.

\begin{figure}[hbt!] 
\centering             
\includegraphics[width=1\textwidth]{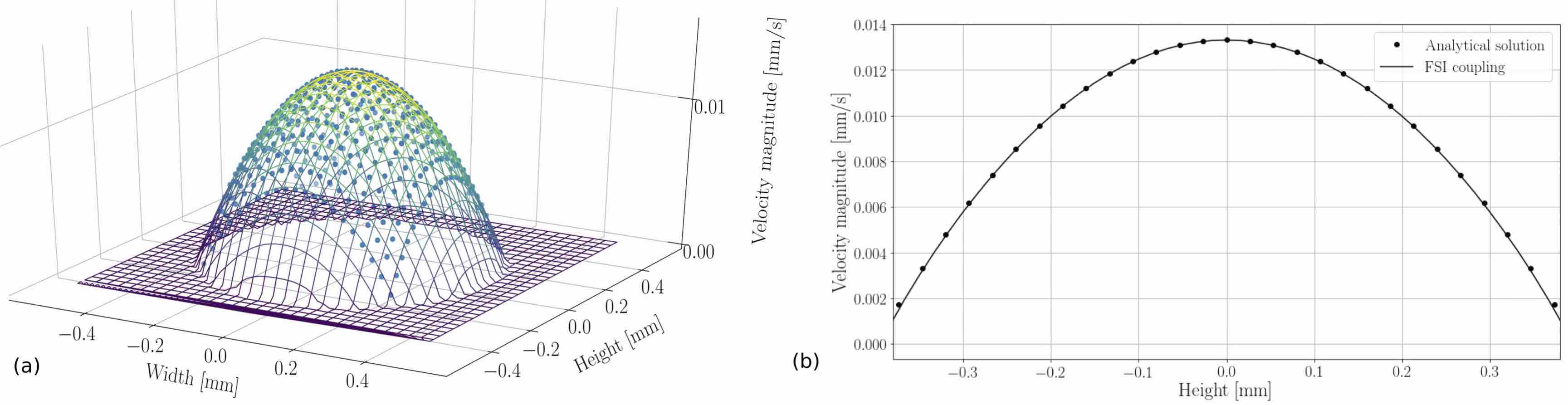}
\caption{Velocity profile of the 3D channel flow benchmark problem for: a)  The immersed boundary approach (colored contour lines) and the analytical solution (blue dots); b) 1D velocity profile from the solutions of the cross-section in a).}
\label{fig:3d_cmp_vel}
\end{figure}

To compare the result with the analytical solution, the velocity profile in the x-plane in 2D is extracted and plotted together with an analytical solution derived from Poiseuille's law in 3D.  Fig.~\ref{fig:3d_cmp_vel}a shows the alignment of the velocity profile of the immersed boundary approach with the analytical solution.  This observation is confirmed, when we compare the 1D velocity profile at y = 0.5~mm with the same analytical solution (Fig.~\ref{fig:3d_cmp_vel}b).

In conclusion of this benchmark section, we observe that the benchmark problems demonstrate the ability of the immersed boundary approach to resolve the boundary between the solid and fluid and to simulate the fluid flow accordingly in two and three dimensions. 

\subsection{Rough Fracture Surfaces} \label{sec:roughFractureSurfaces}
This section investigates fluid flow through rough rock fracture geometries and the respective responses to increasing normal load or fluid injection pressures. 
For this purpose, two numerical experiments are conducted: (1) Fracture closure with increasing confining stresses normal to the fracture and (2) fracture opening with increasing fluid pressures in the fracture. 

\begin{figure}[hbt!] 
\centering             
\includegraphics[width=0.4\textwidth]{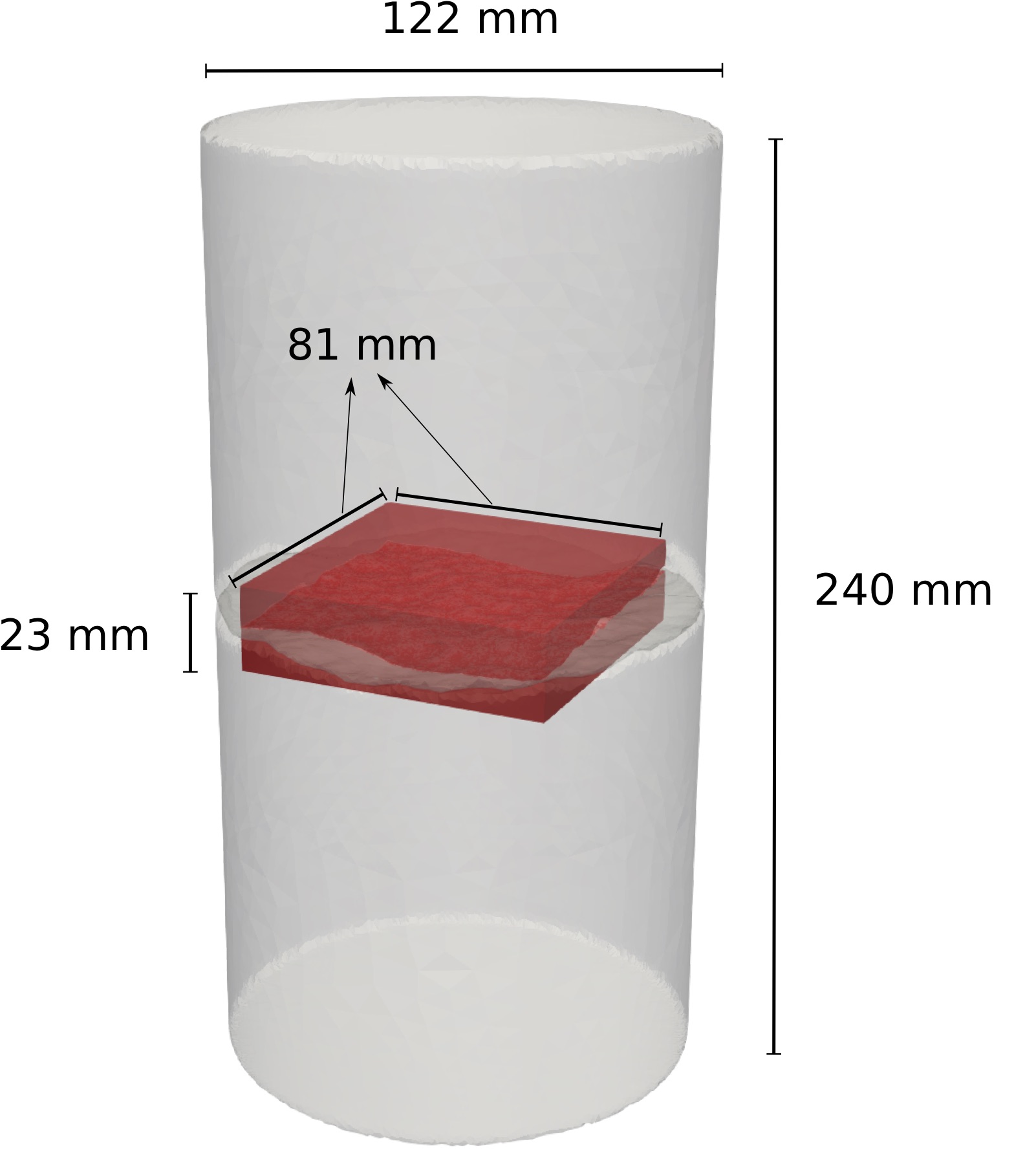}
\caption{Solid geometry in red, extracted from the original cylinder (gray), used for contact simulations in \cite{PVZ+18}.}
\label{fig:rf_geometries}
\end{figure}

We obtain the geometries for the solid domain $\Omega_s^t$ from the contact simulations in \cite{PVZ+18}. There, numerical experiments are conducted on modeled rock fracture geometries under increasing normal loads from 0 to 22~MPa, using linear elasticity and linearized contact conditions (see e.g. \cite{KO88}). The contact conditions between the nonmatching surface meshes were thereby resolved using a mortar approach with dual Lagrange multipliers \cite{WK03,PVN+18a,PVZ+18}. The geometries originate from a granodiorite rock specimen from the Grimsel Test Site located in the Swiss Alps, which were previously used in laboratory experiments \cite{vogler_2016b,vogler_2016c,vogler_2018}.  The specimen is of cylindrical shape, with a fracture normal to the cylinder axis in the center of the specimen.  Fracture topographies were obtained through photogrammetry scans, which allowed the creation of numerical meshes for the upper and lower half of the specimen in \cite{PVZ+18}. 
Because we want to focus on the processes near the fracture, we do not simulate the entire cylinder. Instead, we extract a rectangular cubical region around the fracture (red part in Fig.~\ref{fig:rf_geometries}), in order to immerse the resulting solid in a rectangular structured mesh for the fluid. This helps setting up the fluid problem, as the role of the boundaries and the boundary conditions to be used become clearer. In addition, the resolution in the fracture region is increased, as nodes to model peripheral regions are not needed, as would be the case if  the entire cylinder would be used. 

The original cylinder from the contact simulations by \cite{PVZ+18}, had a diameter of 122~mm and a height of 240~mm.  From this we extract solid geometries with lengths and widths of 81~mm. Depending on the experiment type, the height is either 9~mm or 23~mm.

\subsubsection{Fluid flow in closing fracture}
These simulations model the change of flow patterns in a fracture which is subjected to increasing normal loads.  Here, we use thinner geometries with a height of 9~mm, which are obtained from different stages of the closure process (i.e., confining pressures ranging from 0.25 to 20~MPa), that is, the geometry itself reflects the fracture under different confining pressures, but does not have a normal loading in the HM-simulation, in order to reduce the complexity of the setup.

\begin{figure}[hbt!] 
\centering             
\includegraphics[width=1\textwidth]{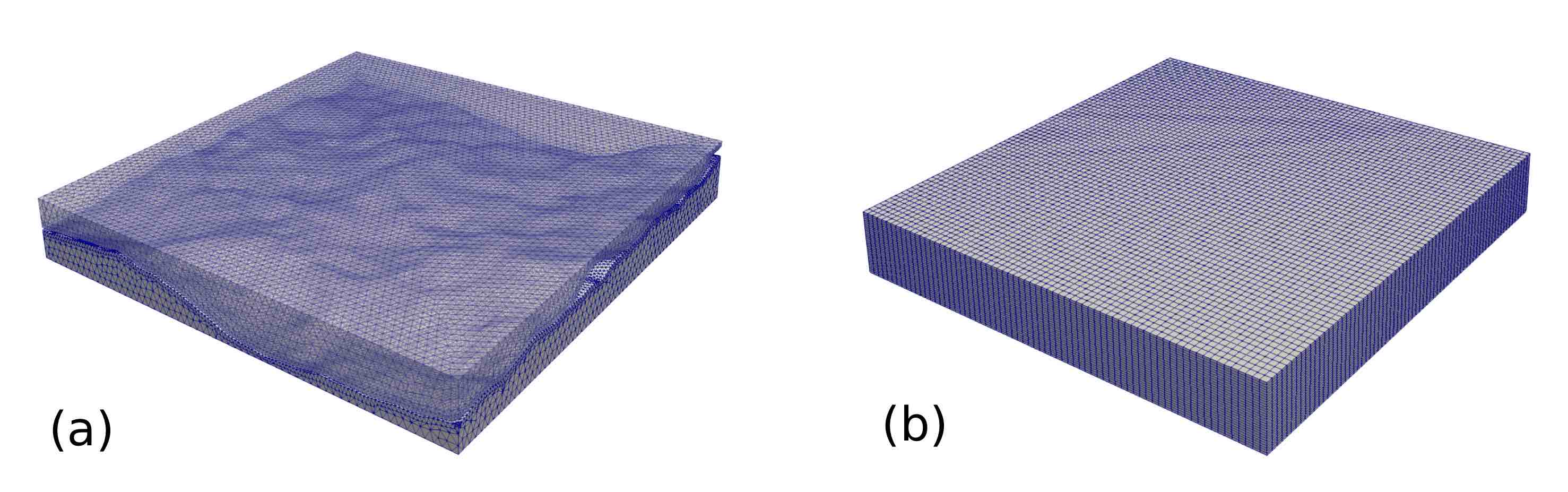}
\caption{Mesh geometries: a) Unstructured solid mesh; and b) Structured fluid mesh.}
\label{fig:RF_Mesh}
\end{figure}

Solid meshes resolution ranges between 100'637 and 123'231 nodes.  The meshes are refined along the fracture surface to achieve a higher resolution at the fluid-solid boundary (Fig.~\ref{fig:RF_Mesh}a). The structured fluid mesh has the same length, width and height as the solid mesh.  It has 1'212'741 nodes and is more refined in the z-direction to gain a better resolution parallel to the fracture plane (Fig.~\ref{fig:RF_Mesh}b). 

\begin{figure}[hbt!] 
\centering             
\includegraphics[width=1\textwidth]{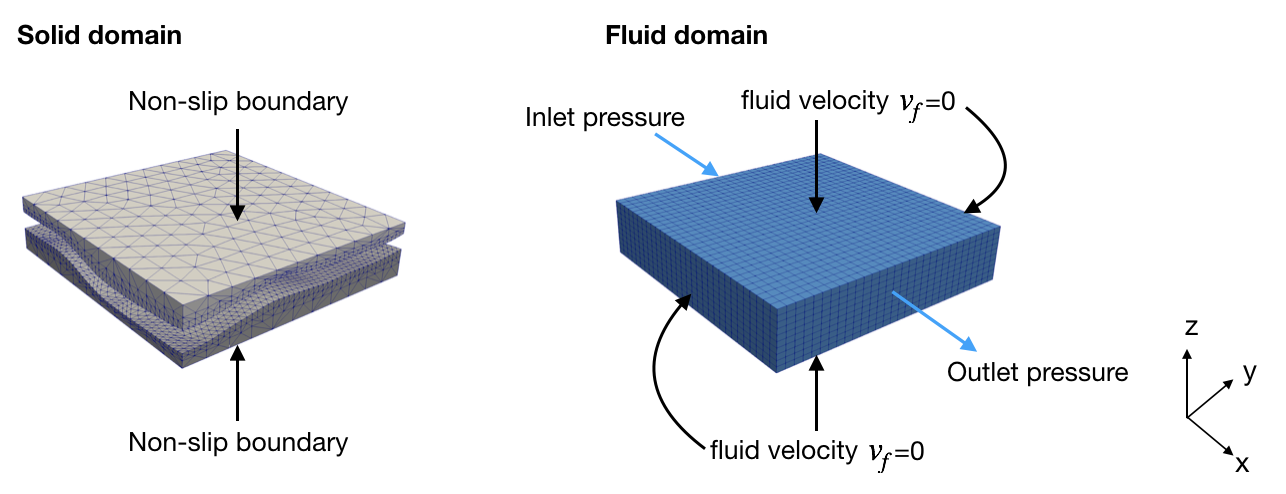}
\caption{Boundary conditions for numerical simulation of fluid flow in the closing fracture.}
\label{fig:RF_bc}
\end{figure}

We apply Dirichlet boundary conditions to the top and bottom boundaries of the solid (Fig.~\ref{fig:RF_bc}a).  On the fluid, we apply non-slip conditions at the top, bottom, front and back boundaries.  We set the fluid pressure on the left (1~MPa) and right side (0~MPa) of the fluid domain (Fig.~\ref{fig:RF_bc}b), thereby establishing a fluid pressure gradient across the fracture.
\begin{figure}[hbt!] 
\centering             
\includegraphics[width=0.7\textwidth]{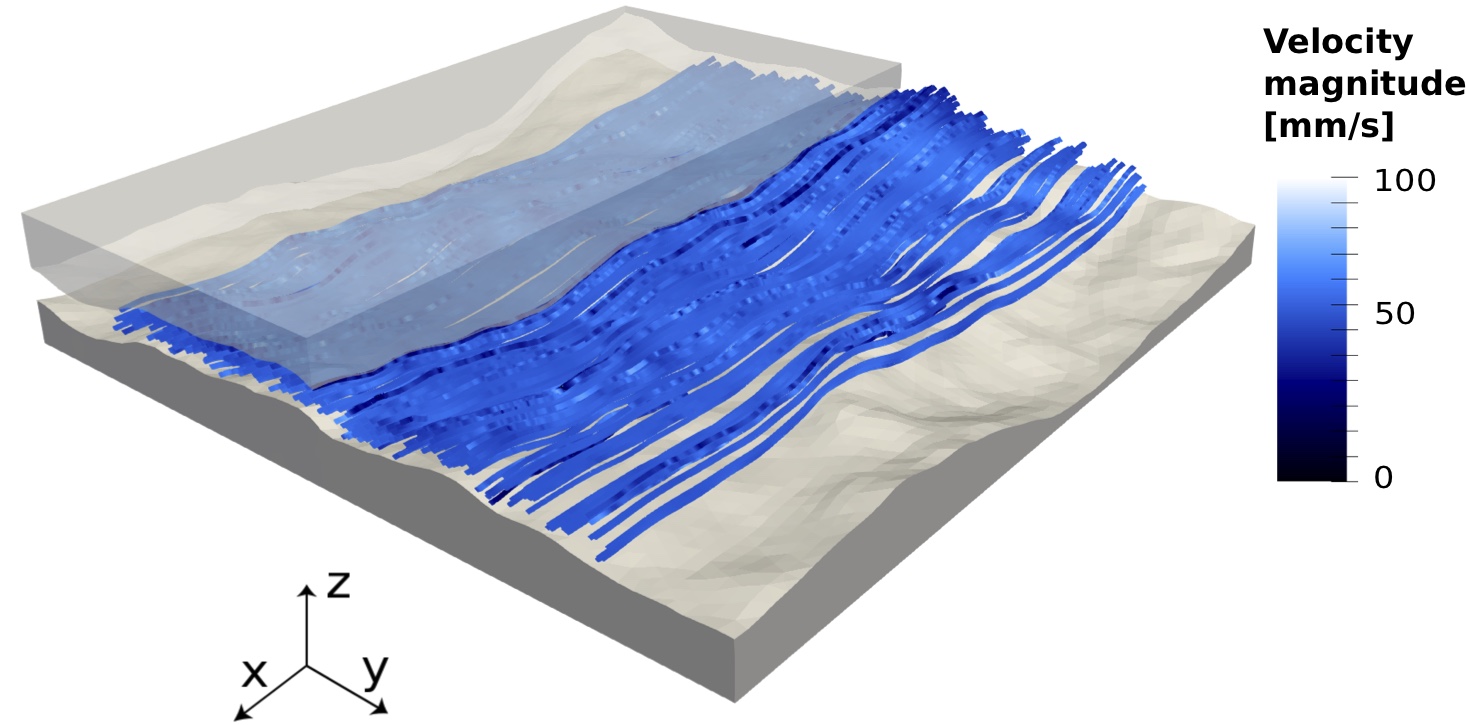}
\caption{Exemplary fluid velocity streamlines in an open fracture.}
\label{fig:RF_Velo_Strm}
\end{figure}

We visualize the setup in Fig.~\ref{fig:RF_Velo_Strm}, where we increase the aperture of the fracture in order to clarify the concept. The picture shows the velocity streamlines of the fluid, originating from a sphere in the center of the open fracture geometry. Fluid flow is tortuous against the rough fracture surface with no intersections between the fluid and the solid.

\begin{figure}[hbt!] 
\centering            
\includegraphics[width=1\textwidth]{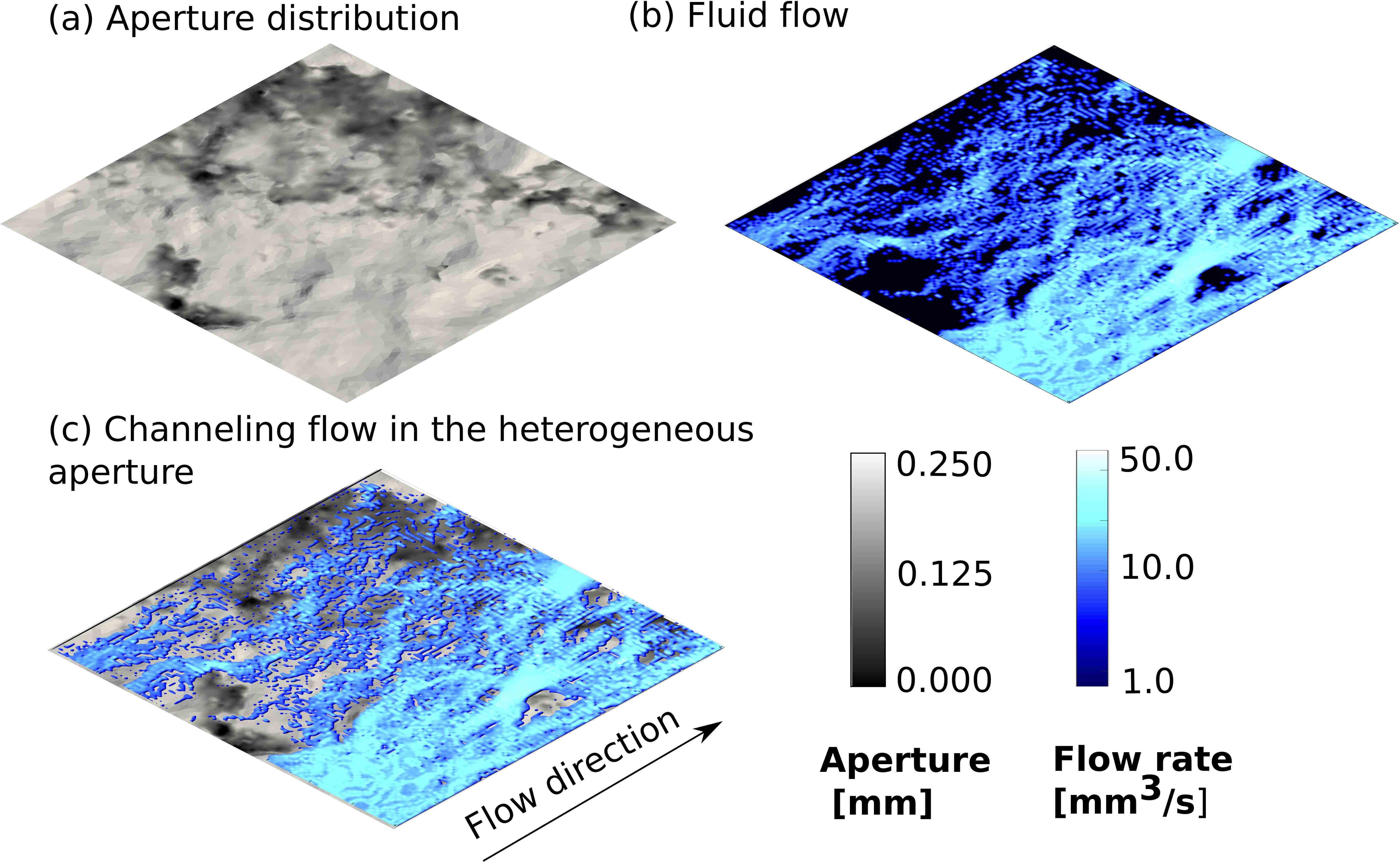}
\caption{Relationship between aperture distributions and the resulting fluid flow field within the fracture: a) Aperture distribution of the fracture for a confining pressure of 0.25~MPa; b) Volumetric fluid flow rates within the fracture, driven by the pressure gradient; c) Composite image, showing the channeling flow paths due to the aperture distribution.}
\label{fig:RF_ChannelFlow}
\end{figure}

Fig.~\ref{fig:RF_ChannelFlow} shows the fluid flow rate averaged in the z-direction (to ease visual comparison) and the aperture distribution at a confining pressure of 0.25~MPa.  The aperture field is strongly heterogeneous  (Fig.~\ref{fig:RF_ChannelFlow}a), resulting in the characteristic preferential fluid flow paths \cite{tsang_1984,brown_1987,tsang_1987,nemoto_2009} that can be seen in Fig.~\ref{fig:RF_ChannelFlow}b.  The two plots are combined in Fig.~\ref{fig:RF_ChannelFlow}c to illustrate the dependence of the fluid flow field on the aperture field more clearly.  As expected, flow channels form in regions with large apertures (white) or around large regions of fracture wall contact (black).

\begin{figure}[hbt!] 
\centering             
\includegraphics[width=1\textwidth]{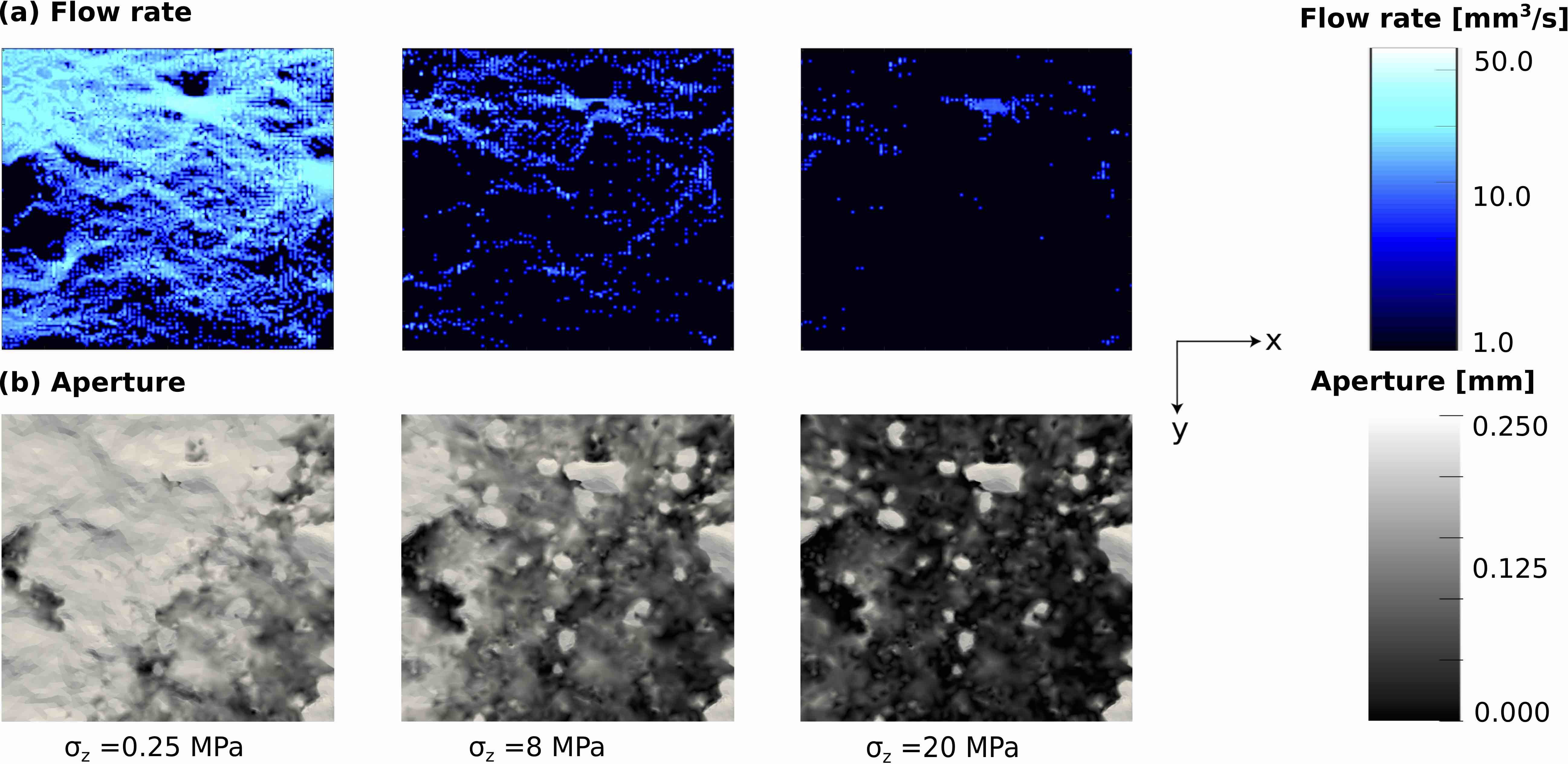}
\caption{Spatial distribution on the fracture surface under different normal stresses for: a) Volumetric fluid flow rates; b) Aperture widths.}
\label{fig:RF_Flowpath}
\end{figure}

Fig.~\ref{fig:RF_Flowpath} depicts the flow path evolution under increasing normal loads. We again average flow rates in the z-direction and view aperture values from above to ease visual comparison.  Here, we show the spatial distribution of fluid flow rates, as in Fig.~\ref{fig:RF_ChannelFlow}b, under normal loads of 0.25, 8 and 10~MPa.  For small loads, the majority of the fracture is essentially open (\ref{fig:RF_Flowpath}b), resulting in fluid flow across large areas of the fracture (\ref{fig:RF_Flowpath}a).  Nonetheless, areas with large aperture widths experience particular high fluid flow rates, while regions, where the fracture walls are already in contact, are void of fluid flow.  With increasing normal load, fracture areas, which were open for small normal loads, are now closed and do not experience significant fluid flow rates. The importance of heterogeneous aperture width fields and their fine resolution is apparent across all loading stages, as large- and small-aperture regions always govern fluid flow fields.

\begin{figure}[hbt!] 
\centering             
\includegraphics[width=1\textwidth]{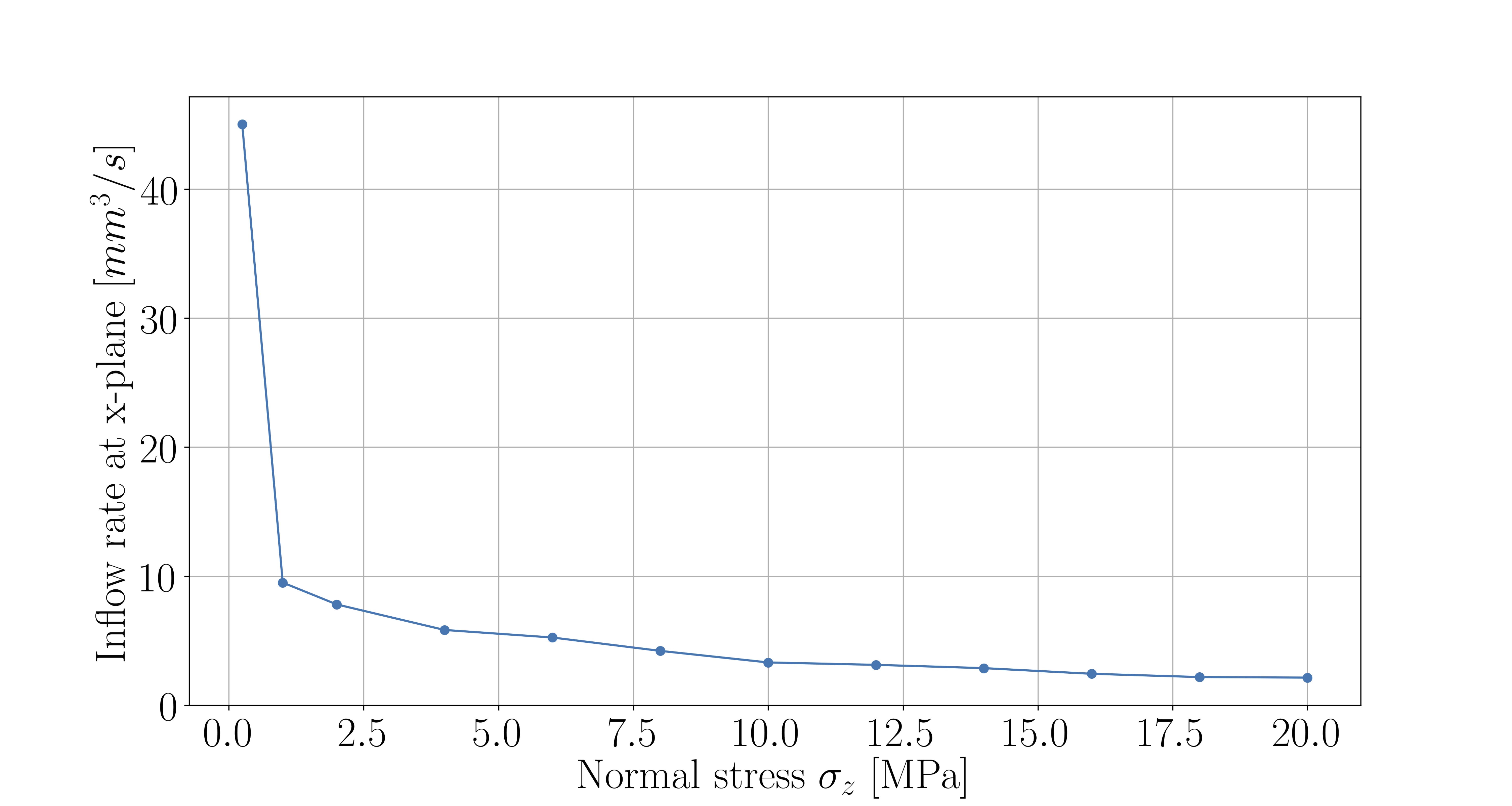}
\caption{Volumetric fluid inflow rate into the fracture through the x-plane, i.e. along the y-plane, versus the confining pressure or normal stress $\sigma_z$ across the fracture.}
\label{fig:RF_sig_fl}
\end{figure}

To illustrate the rapid changes in fracture permeability that a fracture under load experiences, Fig.~\ref{fig:RF_sig_fl} shows the non-linear behavior of the fluid inflow rate into the fracture versus normal load across the fracture \cite{witherspoon_1980,raven_1985,pyraknolte_2000,watanabe_2008,vogler_2016,vogler_2018}.  The inflow rate is calculated by integrating the fluid velocity over the x-plane cross-section at the left side (inflow direction) of the y-plane.
The total fluid flow rate through the fracture changes most drastically at initial normal load increases during small normal loads, where fractures are known to experience the largest aperture closure relative to increases in normal load \cite{bandis_1983,matsuki_2008,zangerl_2008,vogler_2018,kling_2018}.  While fracture closure is more pronounced during early normal loading stages, fluid flow paths through regions of large aperture widths persist even at higher loading stages, as their large-aperture regions may not close entirely (Fig.~\ref{fig:RF_Flowpath}).

\subsubsection{Fracture opening}
These simulations aim to demonstrate fracture opening with the immersed boundary numerical modeling approach.  In particular, we test fracture opening under increasing fluid pressure in the fracture. For this purpose, we compare results for constant fluid pressures in the fracture of 10~and 20~MPa respectively.

A fracture under a normal load of 20~MPa serves as our starting geometry, i.e. in a quasi closed state (see last row in Fig.~\ref{fig:RF_Flowpath}). Compared to  the previous section, the solid mesh is thicker, as the focus in this section is on the mechanical response of the fracture due to changes in fluid pressure. The overall height of the solid is $\sim$23~mm, resulting in more finite element nodes along the z-axis, which helps in resolving the mechanical displacements.

\begin{figure}[hbt!] 
\centering             
\includegraphics[width=1\textwidth]{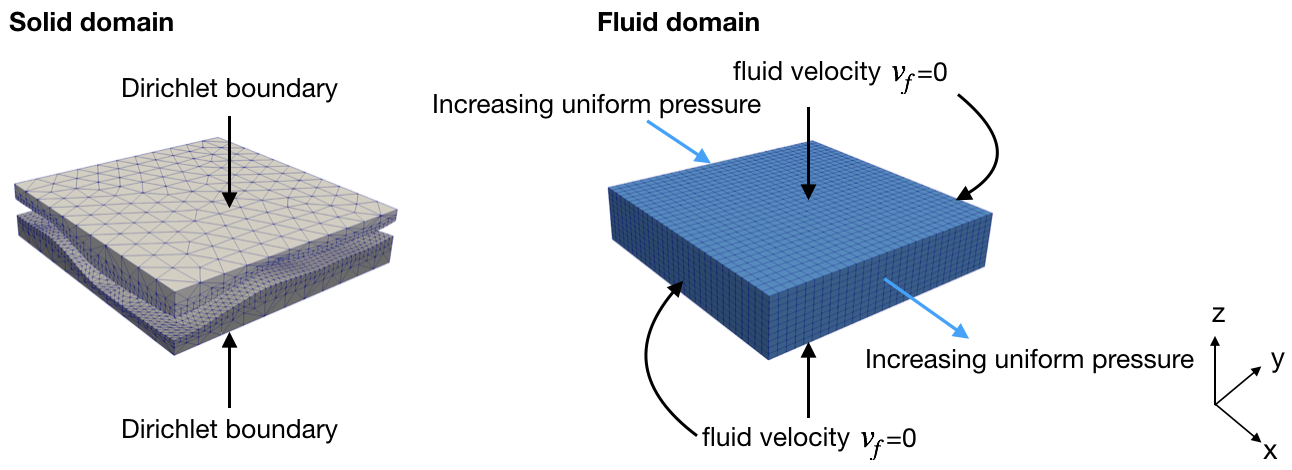}
\caption{Boundary conditions for the solid and fluid domains for the numerical simulations of fracture opening due to increasing fluid pressure.}
\label{fig:FC_bc_fsi_Dir}
\end{figure}

We apply Dirichlet boundary conditions at the top and bottom boundaries of the solid domain and we set a uniform fluid pressure, that is the same fracture inlet pressure and outlet pressure, at the left and right boundaries of the fluid domain (Fig.~\ref{fig:FC_bc_fsi_Dir}). In addition, no-slip boundary conditions are applied at the top, bottom, front and back boundaries of the fluid domain. 

\begin{figure}[hbt!] 
\centering             
\includegraphics[width=1\textwidth]{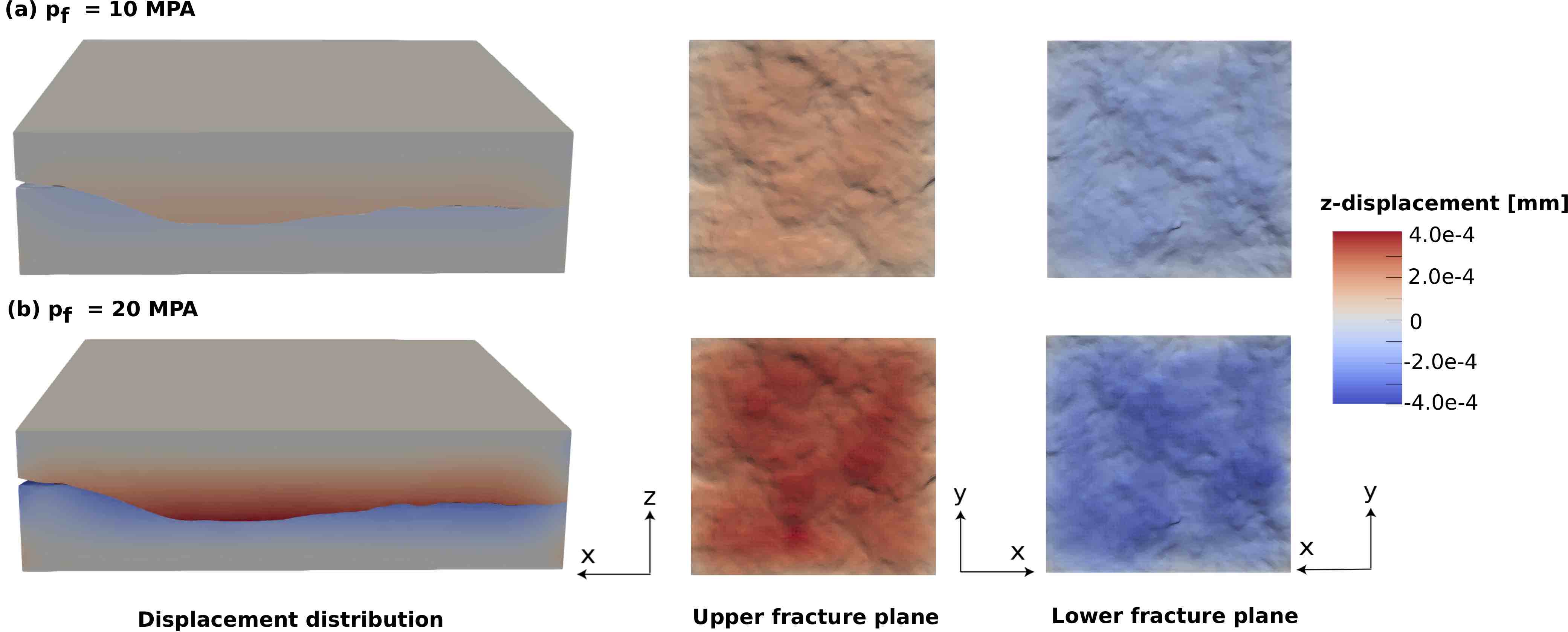}
\caption{Displacement of the solid fracture walls in the z-direction, due to fluid pressure $P_f$ increase to 10 (left column) and 20~MPa (right column). 
The top row shows the fracture geometries from the side, the middle row shows displacements on the top geometry half and the bottom row shows displacements on the bottom geometry half. The fracture planes are displayed as the correspondent Cartesian coordinate axes. 
Blue colors denote displacements in negative z-direction (i.e. downward) and red colors in positive z-direction (i.e. upward). Displacements converge to zero at the top and bottom boundaries of the solid, due to the Dirichlet boundary conditions applied there (Fig.~\ref{fig:FC_bc_fsi_Dir}).}
\label{fig:fo_dc_disp_z}
\end{figure}

Figure \ref{fig:fo_dc_disp_z} shows the response of the fracture to a fluid pressure of 10~MPa (left column) and 20~MPa (right column).  The top row of the figure is a cross-section through the fracture.  Blue colors denote displacements downward and red colors upward.  Due to the Dirichlet boundary conditions, applied to the top and bottom boundaries of the solid domain (Fig.~\ref{fig:FC_bc_fsi_Dir}), displacements converge to zero as these solid boundaries are reached. The middle and bottom row of Figure \ref{fig:fo_dc_disp_z} show displacements in the z-direction for the upper and for the lower fracture surfaces or planes, respectively. This demonstrates, that the immersed boundary approach can simulate the fluid pressure acting on the solid, thereby causing a mechanical response in the solid.
\section{Conclusions}
We presented fully coupled hydro-mechanical (HM) numerical simulations of fluid-saturated, rough-walled fractures. To this end, we adapted an immersed boundary  method. The approach allows us to solve the coupled problem without an explicit representation of the complex boundary between the solid and the fluid.  Instead, the coupling between the two domains relies entirely on the computation of volume integrals.  This immensely simplifies the setup of HM-simulations. Additionally, our implementation solely relies on software components that are designed for parallel computing. 

We validated our approach with two and three dimensional benchmarks and also with geometries from a real fracture under load, which was obtained from a laboratory rock core specimen. The numerical experiments show, that the method replicates Poiseuille flow in three dimensions, that it replicates the change of fluid flow patterns in a fracture under increasing normal loads (confining pressures) across the fracture and that it can also simulate the deformation of the solid when the fluid pressure inside the fracture changes. The latter was demonstrated by increasing the fluid pressure inside the fracture, causing fracture opening.

The methodology enables the computationally efficient investigation of complex fluid flow patterns in rough-walled fractures that exhibit complex fracture surface geometries. The method can also numerically simulate the interaction between the fluid and the solid.  In particular, the ability to simulate the interaction between the solid and the fluid, due to increases in the fluid pressure inside the fracture, can make this method a valuable tool for investigating HM processes, for example, during hydraulic stimulation. 

We employ variational transfer operators to map fluid pressure and velocity between the fluid and the solid domains. The same concept could also be used to extend HM simulations, using staggered approaches and coupling, to include thermal processes.

\acknowledgements{
We gratefully acknowledge funding by the Swiss Competence Center for Energy Research - Supply of Electricity (SCCER-SoE), by Innosuisse - Swiss Innovation Agency under Grant Number 28305.1 and the Swiss Federal Office of Energy (SFOE) under Grant Number SI/500676-02. The Werner Siemens-Stiftung (Foundation) is thanked for its support of the Geothermal Energy and Geofluids group.
}


\bibliographystyle{plain} 
\bibliography{bibliography}

\end{document}